\documentclass[12pt,a4paper]{article}
\usepackage{epsfig}
\usepackage{color}
\usepackage{amsmath}
\usepackage{amssymb}
\usepackage{lscape}

\makeatletter
\@addtoreset{equation}{section}

\makeatother

\begin{document}

%\fontsize{14pt}{0.8cm}\selectfont
\fontsize{11pt}{0.5cm}\selectfont

\title{
\vbox{
\baselineskip 14pt
\hfill \hbox{\normalsize KEK-TH-2358
}} \vskip 1cm
\bf \Large  
Gauge Symmetry Restoration by Higgs Condensation   \\
  in Flux Compactifications on Coset Spaces
\vskip .5cm
}

\author{
Satoshi Iso$^{a,b}$  \thanks{E-mail: \tt iso(at)post.kek.jp}, 
Noriaki Kitazawa$^{c}$ \thanks{E-mail: \tt noriaki.kitazawa(at)tmu.ac.jp},
Takao Suyama$^{a}$ \thanks{E-mail: \tt tsuyama(at)post.kek.jp}
\bigskip\\
\it \normalsize
$^a$ Theory Center, High Energy Accelerator Research Organization (KEK), \\
\it  \normalsize   Tsukuba, Ibaraki 305-0801, Japan \\
\it  \normalsize
$^b$Graduate University for Advanced Studies (SOKENDAI),\\
\it \normalsize Tsukuba, Ibaraki 305-0801, Japan \\
\it  \normalsize 
$^c$Department of Physics, Tokyo Metropolitan University,\\
\it  \normalsize Hachioji, Tokyo 192-0397, Japan \\
\smallskip
}
\date{\today}

\maketitle

\abstract{\normalsize
Extra-dimensional components of gauge fields in higher-dimensional gauge theories
 will play a role of the Higgs field and become tachyonic after Kaluza-Klein compactifications
 on internal spaces with (topologically nontrivial) gauge field backgrounds.
Its condensation is then expected to  break  gauge symmetries spontaneously. 
But, contrary to the expectation, 
 some models exhibit restoration of gauge symmetries. 
In this paper, by considering  all the massive Kaluza-Klein excitations of gauge fields,
 we explicitly show that some of them indeed become massless 
 at the minimum of the Higgs potential and 
 restore (a part of) the gauge symmetries which are broken by gauge field backgrounds.
We particularly consider compactifications on 
 $S^2$ with monopole-like fluxes and also on
 $\mathbb{CP}^2$ with instanton and monopole-like fluxes.
In some cases, the gauge symmetry is fully restored, as argued in previous literatures.
In other cases,
 there is a stable vacuum  with a partial restoration of the gauge symmetry after Higgs condensation.
Topological structure of the gauge field configurations prevent the gauge symmetries to be restored.}

\vspace{1cm}

\tableofcontents

\newpage

\section{Introduction}

\vspace{5mm}

The dynamics of gauge symmetry breaking is yet to be investigated,
 especially when it is caused by the elementary Higgs scalar field with a non-trivial potential.
The mechanism of gauge symmetry breaking or the origin of 
the Higgs potential is highly required.
Among numerous proposals or models including radiative symmetry breaking mechanism
  (see e.g.,
\cite{Coleman:1973jx,Ibanez:1982fr,Inoue:1982pi,Bardeen:1995kv,Iso:2012jn,Iso:2009nw,
 Antoniadis:2000tq})
 and extra dimensions
 (see e.g.,
 \cite{Manton:1979kb,Fairlie:1979zy,Fairlie:1979at,Hosotani:1983xw,Hosotani:1983vn,Hosotani:1988bm,
 Antoniadis:1990ew,Antoniadis:1998ig,Hatanaka:1998yp,Arkani-Hamed:1998sfv,Randall:1999ee,
 Randall:1999vf,Dvali:2001qr,Arkani-Hamed:2001nha,Lim:2006bx,
 Kitazawa:2012hr,Iso:2015mva}),
a possibility to understand the origin of Higgs potential in 
the context of higher dimensional gauge theory has been widely investigated.
In the present paper, we revisit the gauge symmetry breaking by the 
coset space dimensional reductions of higher-dimensional gauge theories
 with background gauge fluxes (see \cite{Kapetanakis:1992hf} for review,
 and see e.g., \cite{Andriot:2020ola} for dimensional reduction to non-coset spaces).
The basic idea of this construction appeared in \cite{Manton:1979kb} which realizes the bosonic part of the Weinberg-Salam model from the six-dimensional Yang-Mills theory. 
In this construction, the Higgs potential of the double-well type dynamically appears, and
this class of models 
 have been applied to the gauge-Higgs unification models of the electroweak theory
 \cite{Dvali:2001qr}.

Background gauge fluxes in  compact spaces 
in higher-dimensional gauge theories
are originally introduced to stabilize the compact space  
 \cite{Randjbar-Daemi:1983llh,Randjbar-Daemi:1983zks,Randjbar-Daemi:1982opc,
 Randjbar-Daemi:1983xth,Schellekens:1984dm,Schellekens:1985ks}
 in the context of the Einstein-Yang-Mills theory, 
 and further  developed in the studies of flux compactifications in string theories
 (for reviews, see \cite{Douglas:2006es,Blumenhagen:2006ci,Grana:2005jc} ). 
 The well studied examples of compact spaces are coset spaces $G/H$, such as
 $S^2=$SU$(2)/$U$(1)$ or $\mathbb{CP}^2=$SU$(3)/($SU$(2)\times$U$(1)$),
 and (in)stability of such compactifications in the presence of  gravity have been extensively investigated.
If tachyonic fields appear, the solution becomes unstable and 
their condensations will generate a new vacuum solution. 
In particular, such tachyonic fields can be utilized as candidates  of the Higgs scalars, 
and understanding of the shape of tachyon potential and the pattern of gauge symmetry breaking
is an important issue to be investigated. 

In flux compactifications,  
the original gauge symmetry in higher-dimensions 
 is explicitly broken by the background gauge fluxes in the compact spaces, 
 and the Higgs vacuum expectation value is expected to further  break 
 some part of the remaining gauge symmetries spontaneously in four-dimensional effective theory. 
In previous literatures, most studies have been focussed on low lying states in the 
effective four-dimensional theories after compactifications. 
Among an infinitely many fields,
only massless fields are usually taken into considerations in the effective theory, 
 and all the other higher excited modes are neglected. 
It is justified when we consider low-energy physics below the scale of the compact spaces, 
but when we investigate the condensation of the tachyonic field, 
the massive modes will also play an important role since 
the mass scale in the Higgs potential is typically the same as masses of Kaluza-Klein higher modes. 
Especially,  when the Higgs acquires vacuum expectation value,
we need to take care of a possibility that
 some of the massive Kaluza-Klein modes may become massless. 

In this paper, 
 we investigate dynamics of the Higgs condensation in several simple models
on coset spaces, such as $S^2=$SU$(2)/$U$(1)$ and $\mathbb{CP}^2=$SU$(3)/($SU$(2)\times$U$(1)$),
with all the massive Kaluza-Klein modes included. 
We find that,   
 although the Higgs vacuum expectation value itself breaks a part of the remaining gauge symmetries
 and the corresponding gauge bosons indeed become massive,
 some of the  massive Kaluza-Klein modes will become massless
 and  gauge symmetries are recovered in the four-dimensional effective theory.
In fact, such a possibility was  pointed out in \cite{Palla:1983re,Kozimirov:1988aw}. 
In this paper, we develop a group-theoretic technique which enables us to clarify explicitly
 which Kaluza-Klein modes become massless vectors after the Higgs condensation.
In string theory, it is known that similar gauge symmetry enhancement occurs for condensation of massless scalars, or moduli. 
Non-supersymmetric string theories are discussed in this context recently in \cite{Itoyama:2021itj} and references therein. 

In section \ref{sec:KKcoset}, 
we give a general formulation of the Kaluza-Klein reduction on coset spaces $G/H$
with a topologically non-trivial background gauge field configuration. 
In section \ref{sec:symmetric Higgs},
 we  introduce the notion of ``symmetric field'' \cite{Forgacs:1979zs}
 which corresponds to the zero mode, or a constant mode on 
 the flat compact space without flux.
Interestingly,
 some of these symmetric fields may have non-trivial potentials with a negative mass squared at the origin, 
 and we call them {\it symmetric Higgs fields}.
In section \ref{condensation  S^2},
 we investigate the Higgs condensation in gauge field theories compactified on $S^2$.
We study three different types of models, whose background monopole-like fluxes are different. 
We particularly investigate patterns of gauge symmetry breaking
 when the symmetric Higgs fields have vacuum expectation value. 
In section \ref{condensation CP^2}
 we generalize the analysis on $S^2$ to  $\mathbb{CP}^2$ coset models.
 In this case, since the coset space is $\rm SU(3)/(SU(2)\times U(1))$,
 both of instanton and monopole-like background configurations can exist. 
In one of the examples we study, all the gauge symmetries are restored by the Higgs condensation,
 which cancels the background gauge flux
 as was pointed out in \cite{Palla:1983re,Kozimirov:1988aw}.
There exists another type of models,
 in which a topologically nontrivial gauge field fluxes prevent the gauge symmetries to be recovered,
 and a stable vacuum with a partial restoration of gauge symmetries is realized. 
In the last section we summarize our results and conclude.

There are several Apendices which review various materials necessary for our investigations.
In Appendix \ref{coset space appendix},
 we review the basics of coset space $G/H$,
 and describe $G$ as a principal $H$-bundle in Appendix \ref{principal bundle appendix}.
In Appendix \ref{Maurer-Cartan appendix},
 we review the construction of the background gauge field and the vielbein on $G/H$
 which are provided by the Maurer-Cartan 1-form on $G$. 
In Appendix \ref{EOM appendix},
 we review a proof that the background gauge field satisfies the equations of motion. 
In Appendix \ref{app: explicit formulas},
 concrete forms of the background gauge field
 and the vielbein are given in the case of $S^2={\rm SU}(2)/{\rm U}(1)$. 
In Appendix \ref{mode functions appendix},
 we review a construction of mode expansions on $G/H$
 by using the Peter-Weyl theorem for the mode expansions on $G$.  
In Appendix \ref{app: laplacian and mass formula},
 we explain eigenvalues of the Laplacian on mode functions and mass formula of various fields on $G/H$.
We also show that the symmetric Higgs field has a negative mass squared and becomes tachyonic. 
In Appendix \ref{app: symmetric Higgs field},
 we prove that the symmetric Higgs field satisfies the condition of the symmetric field on $G/H$.

%%%%%%%%%%%%%%%%%%%%%%%%%%%%%%%%%%%%%%%%%%%%
%%%%%%%%%%%%%%%%%%%%%%%%%%%%%%%%%%%%%%%%%%%%

\vspace{1cm}

\section{Kaluza-Klein reduction on coset spaces}
\label{sec:KKcoset}
\vspace{5mm}

\subsection{Action in background gauge fields}
We consider Yang-Mills theory on a $(4+d)$-dimensional manifold  $\mathbb{R}^4\times{\cal M}$ with 
the action 
\begin{equation}
S\ =\ \int dv\,{\rm Tr}\left[ -\frac14F_{MN}F^{MN} \right], \hspace{1cm} dv\ :=\ \frac1{g_{\rm YM}^2} d^{4+d}X\sqrt{-G}, 
\end{equation}
where $M,N=0,1,\cdots,3+d$ and $G_{MN}$ is a metric on $\mathbb{R}^4\times{\cal M}$. 
The overall normalization of the action is chosen such that each matrix component of the gauge field $A_M$ is canonically normalized. 
Our convention for the field strength is 
\begin{equation}
F_{MN}\ :=\ \nabla_MA_N-\nabla_NA_M+i[A_M,A_N], 
\end{equation}
where $\nabla_M$ is the covariant derivative with respect to the metric $G_{MN}$. 

We investigate this theory around a background gauge field $\bar{A}_M$. 
The gauge field $A_M$ is then decomposed as $A_M=\bar{A}_M+a_M$. 
In the following, we often use the notation 
\begin{equation}
\bar{D}_Ma_N\ :=\ \nabla_Ma_N+i[\bar{A}_M,a_N]. 
\end{equation}
We employ the background field gauge 
\begin{equation}
\bar{D}^Ma_M\ =\ \nabla^Ma_M+i[\bar{A}^M,a_M]\ =\ 0. 
\end{equation}
The corresponding gauge-fixing term is given by
\begin{equation}
S_{\rm gf}\ =\ \int dv\,{\rm Tr}\left[ -\frac12\left( \bar{D}^Ma_M \right)^2 \right]. 
\end{equation}
By expanding $a_M$ into Kaluza-Klein modes on ${\cal M}$, 
we can obtain a four-dimensional gauge theory coupled to various matter fields.  

%\vspace{5mm}

Let $x^\mu$ ($\mu=0,\cdots,3$) be coordinates on $\mathbb{R}^4$, and let $y^\alpha$ ($\alpha=1,\cdots,d$) be coordinates on $\cal M$. 
Accordingly,  the gauge field $a_M$ is decomposed into $a_\mu$ and $\phi_\alpha$. 
We assume that the background gauge field $\bar{A}_M$ is of the form 
\begin{equation}
\bar{A}_M\ =\ (0,\bar{A}_\alpha), \hspace{1cm} \partial_\mu \bar{A}_\alpha\ =\ 0. 
\end{equation}
This means that we put an  $x^\mu$-independent gauge flux on ${\cal M}$.
Note that the extra-dimensional components of the gauge field, 
 $\phi_\alpha$, provide a set of adjoint matters, transforming homogeneously under the gauge transformations
 as  they are defined by a difference of two gauge fields $A_\alpha$ and $\bar{A}_\alpha$. 
We also set the background metric of $\mathbb{R}^4\times{\cal M}$  as
\begin{equation}
G_{MN}\ =\ \left[ 
\begin{array}{cc}
\eta_{\mu\nu} & 0 \\
0 & h_{\alpha\beta}(y)
\end{array}
\right]. 
\end{equation}

The total action $S+S_{\rm gf}$ consists of the following three parts:
\begin{eqnarray}
S_1 &=& \int dv\,{\rm Tr}\left[ -\frac14F_{\mu\nu}F^{\mu\nu}-\frac12\left( \partial^\mu a_\mu \right)^2 \right]
\nonumber \\
&=& \int dv\,{\rm Tr}\left[ -\frac12\left( \partial_\mu a_\nu \right)^2-i\partial_\mu a_\nu[a^\mu,a^\nu]+\frac14[a_\mu,a_\nu]^2 \right], 
\end{eqnarray}
\begin{eqnarray}
S_2 &=& \int dv\,{\rm Tr}\left[ -\frac14F_{\mu\alpha}F^{\mu\alpha}-\partial^\mu a_\mu\bar{D}^\alpha\phi_\alpha \right]
\nonumber \\
&=& \int dv\,{\rm Tr}\left[ -\frac12\left( D_\mu\phi_\alpha \right)^2-\frac12\left( \bar{D}_\alpha a_\mu \right)^2+i[a_\mu,\phi_\alpha]\bar{D}^\alpha a^\mu \right], 
\end{eqnarray}
where 
\begin{equation}
D_\mu\phi_\alpha\ :=\ \partial_\mu\phi_\alpha+i[a_\mu,\phi_\alpha], 
\end{equation}
and 
\begin{eqnarray}
S_3 & =&\int dv\,{\rm Tr}\left[ -\frac14F_{\alpha\beta}F^{\alpha\beta}-\frac12\left( \bar{D}^\alpha\phi_\alpha \right)^2 \right]
\nonumber \\
&=& \int dv\,{\rm Tr}\left[ -\frac14\left( \bar{F}_{\alpha\beta}+\bar{D}_\alpha\phi_\beta-\bar{D}_\beta\phi_\alpha+i[\phi_\alpha,\phi_\beta] \right)^2-\frac12\left( \bar{D}^\alpha\phi_\alpha \right)^2 \right], \nonumber \\
   \label{scalar potential original}
\end{eqnarray}
where $\bar{F}_{\alpha\beta}$ is the background field strength of the gauge potential $\bar{A}_\alpha$. 

%\vspace{5mm}

In the Kaluza-Klein reduction, the terms (\ref{scalar potential original}) in $S_3$ give the scalar potential $V(\phi)$
after an integration on $\cal M$. 
In particular, the mass terms of the scalars around $\phi_\alpha=0$ come from the following terms
\begin{equation}
{\rm Tr}\left[ \frac12\left( \bar{D}_\alpha\phi_\beta \right)^2-\frac12\phi^\alpha R_{\alpha\beta}\phi^\beta-i\phi^\alpha [\bar{F}_{\alpha\beta},\phi^\beta] \right], 
\label{massterms}
\end{equation}
where $R_{\alpha\beta}$ is the Ricci tensor for $h_{\alpha\beta}$ on $\cal{M}$. 
Note that we have used the equations of motion for $\bar{A}_\alpha$ in deriving (\ref{massterms}). 
On the other hand, the mass terms of the vector fields are provided from the terms, 
\begin{equation}
{\rm Tr}\left[ \frac12\left( \bar{D}_\alpha a_\mu+i[\phi_\alpha,a_\mu] \right)^2 \right]. 
   \label{mass term vector}
\end{equation}
The second term gives additional contributions to mass  at $\langle\phi_\alpha\rangle \neq 0$, whose effects
 we will investigate in sections \ref{condensation S^2} for ${\cal M}=S^2$
 and section \ref{condensation CP^2} for ${\cal M}=\mathbb{CP}^2$. 
 We show that some of the massive Kaluza-Klein modes become massless by the second term. 

\vspace{5mm}

%%%%%%%%%%%%%%%%%%%%%%%%%%%%%%%%%%%%%%%%%%%
\subsection{Coset space $G/H$}
In the following, we focus our attention on a compactification on a coset space. 
See e.g. \cite{Salam:1981xd, Kapetanakis:1992hf} for more details.

We consider a coset space ${\cal M}=G/H$ where 
$G$ and $H$ are Lie groups with $H\subset G$. 
Let us decompose generators of $G$ as $(\{t_a\}, \{t_m\})$ where $\{t_a\}$  ($a=1, \cdots \dim H$) are a set of generators of $H$.
Note that $\{t_m\}$ ($m=1, \cdots, d=\dim G -\dim H)$
correspond to a basis of the tangent space of $G/H$. 
%, so that the index $m$ runs from 1 to $d$. 
We assume that the generators $t_a, t_m$ satisfy the following commutation relations 
\begin{equation}
[t_a,t_b]\ =\ if^c{}_{ab}t_c, \hspace{1cm} [t_a,t_m]\ =\ if^n{}_{am}t_n, \hspace{1cm} [t_m,t_n]\ =\ if^a{}_{mn}t_a. 
   \label{symmetric coset commutators}
\end{equation}
A coset space whose generators satisfy the commutation relations of this form is said to be symmetric. 
In the following, we use $a,b,c$ for generators of $H$ and $m,n$ for generators along $G/H$.
The indices $m, n$ also represent those of coordinates of the tangent space on ${\cal M}=G/H$.
 In this paper, we discuss two examples of symmetric coset spaces, namely $S^2=SU(2)/U(1)$ and 
$\mathbb{CP}^2=SU(3)/U(2)$. 
In these cases, $t_a$ are represented in terms of block-diagonal matrices, while $t_m$ are given in terms of block-off-diagonal matrices,
and their commutation relations are apparently of the form (\ref{symmetric coset commutators}). 
Non-symmetric coset spaces are discussed in \cite{Forgacs:1985vp}. 

%%%%%%%%%%%%%%%%%%%%%%%%%%%%%%%
\subsection{Metric and background gauge field on $G/H$ }   \label{{Metric and background gauge field on $G/H$ }}
For a given coset space $G/H$, there is a ``natural'' choice for the vielbein $e^m_\alpha$ and the background gauge field $\bar{A}_\alpha$. 
Suppose we have a local embedding $g: y^\alpha \in G/H \rightarrow g(y) \in G$.
Then the Maurer-Cartan 1-form $g^{-1}dg$ {restricted on $g(G/H)$
is written as a sum 
\begin{equation}
g^{-1}dg = i e^a(y) t_a + i e^m(y) t_m,
\end{equation}
 where $e^m$ gives the natural choice of the vielbein on $G/H$
 while $e^a$ provides the gauge field on the coset space. 
Indeed, $A=e^a t_a$ transforms under $g \rightarrow gh$ for $h \in H$,
 which is a gauge transformation as explained in Appendix \ref{Maurer-Cartan appendix},
 as $A \rightarrow h^{-1}A h - ih^{-1} dh$. 
In the following,
 we will consider gauge group $G_{\rm YM}$ that includes $H$ as $H \subset G_{\rm YM}$
 and define the background gauge field on the coset space by 
\begin{equation}
\bar{A} =\bar{A}_\alpha\ dy^{\alpha}:=\ e^a_\alpha (y) T_a dy^{\alpha}, 
   \label{background gauge field}
\end{equation}
 where $T_a$ are generators of the gauge group $G_{\rm YM}$
 which are the corresponding embedding of the generators $t_a$ of $H$ into the Lie algebra 
 $\mathfrak{g}_{\rm YM}$ of $G_{\rm YM}$. 
In this paper,
 we consider various different embeddings of $H$ into $G_{\rm YM}$ for $G/H=S^2$ and $\mathbb{CP}^2$.

%\vspace{5mm}

Interestingly, this background gauge field $\bar{A}_\alpha$ automatically satisfies the equations of motion 
\begin{equation}
\bar{D}^\alpha\bar{F}_{\alpha\beta}\ =\ \nabla^\alpha\bar{F}_{\alpha\beta}+i[\bar{A}^\alpha,\bar{F}_{\alpha\beta}] \ =\ 0
   \label{eom}
\end{equation}
with respect to the vielbein $e^m_\alpha$ \cite{Randjbar-Daemi:1982bjy}. 
This can be checked as follows. 
First, the spin connection $\omega_\alpha{}^m{}_n$
 defined by $de^m=- \omega^m{}_n \wedge e^n$,
 is obtained from the relation 
\begin{equation}
d(g^{-1}dg)=-g^{-1}dg \wedge g^{-1}dg
\end{equation}
or equivalently, from the relation (\ref{structure eqs 2}) 
as 
\begin{equation}
\omega_\alpha{}^m{}_n\ =\ -f^m{}_{an}e^a_\alpha.
   \label{spin connection}
\end{equation}
Thus,
 the spin connection is written in terms of the component of the background gauge field $e^a_\alpha$, 
 and the covariant derivative $\nabla^\alpha \bar{F}_{\alpha\beta}$ with respect to the metric on $G/H$
 has the same form as the second term  in (\ref{eom}). 
Second, by using the equation (\ref{structure eqs 1}), 
the field strength of $\bar{A}_\alpha$ turns out to be 
\begin{equation}
\bar{F}_{\alpha\beta}\ =\ e^m_\alpha e^n_\beta f^a{}_{mn}T_a.
\label{fieldstrength}
\end{equation}
Thus, the gauge field strength is non-vanishing in
the $H$ subgroup of $G_{\rm YM}$. 
Inserting these expressions of  (\ref{spin connection}) and (\ref{fieldstrength}) 
into (\ref{eom}), we find that it reduces to the Jacobi identity for the structure constants
and  the background gauge field indeed satisfies the equations of motion. 
See Appendix \ref{EOM appendix} for more details.

\vspace{5mm}
%%%%%%%%%%%%%%%%%%%%%%%%%%%%%%%%
\subsection{Covariant derivative on $G/H$}
Since the spin connection $\omega_\alpha{}^m{}_n$ and the background gauge field $\bar{A}_\alpha$ are given
 in terms of the same quantity $e^a_\alpha$, the covariant derivative of $\phi_m:=e^\alpha_m\phi_\alpha$ can be written as
 \begin{equation}
\bar{D}_\alpha\phi_m\ := {\partial_\alpha\phi_m- \omega_\alpha{}^n{}_m \phi_n + i[\bar{A}_\alpha, \phi_m]}
=\ \partial_\alpha\phi_m+ie^a_\alpha\left( -if^n{}_{am}\phi_n+[T_a,\phi_m] \right). 
   \label{covariant derivative general}
\end{equation}
This shows that the field $\phi_m$ can be regarded as a field on the flat $\mathbb{R}^d$
 which couples to a gauge field $e^a_\alpha$ as a tensor product of two representations. 
Actually, the second commutation relation in (\ref{symmetric coset commutators})
 implies that $t_m$ form a representation $R_t$ of $H$ on which the generators are given by $if^n{}_{am}$. 
Therefore,
 $\phi_m$ belongs to the tensor product representation of $R_t$ and the adjoint representation of $G_{\rm YM}$ 
and can be decomposed into various irreducible representations of $H$. 
This property plays an important role in the investigations of mass spectrum of various fields with 
different spins and charges on $G/H$. 

%\vspace{5mm}

Besides the beautiful properties we have seen above, 
there are further advantages in choosing a symmetric coset space $G/H$ as the internal manifold $\cal M$. 
Most importantly, many  properties of the complete set of functions on $G/H$ are well-known and 
we can explicitly perform the Kaluza-Klein reduction of any field on $G/H$ \cite{Salam:1981xd}. 
For the coset space $S^2$, these functions  are given by the monopole harmonics \cite{Wu:1976ge}. 
For a general coset space $G/H$, the Peter-Weyl theorem tells us that each mode functions in the complete set on $G/H$ is labeled by a representation of $G$. 
As mentioned above, the field $\phi_m$ on $G/H$ can be regarded as belonging to a particular representation of $H$. 
This information can be incorporated by taking into account the irreducible decomposition of the representation of $G$ with respect to $H$. 
See Appendix \ref{mode functions appendix}  for more details. 

By using the mode functions, the mass of each Kaluza-Klein mode in the four-dimensional sense can be obtained explicitly \cite{Palla:1983re,Schellekens:1984dm}. 
As the mode functions are labeled by the representation of $G$ and its decomposition with respect to $H$, 
the mass is given in terms of group-theoretic quantities. 
Namely, it is given in terms the second Casimir invariants of certain representations. 
We review it in Appendix \ref{app: laplacian and mass formula}, 
 which will be used in the proof of the tachyonic behavior of the symmetric Higgs field
 observed in the following sections. 

%%%%%%%%%%%%%%%%%%%%%%%%%%%%%%%%%%%%%%%%%%%%%%%%%%%%%%%%
%%%%%%%%%%%%%%%%%%%%%%%%%%%%%%%%%%%%%%%%%%%%
%%%%%%%%%%%%%%%%%%%%%%%%%%%%%%%%%%%%%%%%%%%%

\vspace{1cm}

\section{Symmetric Higgs fields}
\label{sec:symmetric Higgs}

\vspace{5mm}

The Kaluza-Klein reduction of a higher-dimensional Yang-Mills theory on $\cal M$ contains infinitely many fields. 
If one wants to employ this theory for phenomenological model buildings, 
it is natural to truncate the theory
 so that the resulting theory contains only a finite number of light fields.
If $\cal M$ is a torus, for example,
 the lowest mass state is given by a constant mode on $\cal M$ for a scalar field. 

When $\cal M$ is a coset space, symmetric fields defined below will 
provide such lowest mass states \cite{Kapetanakis:1992hf} on  ${\cal M}=G/H$. 
A field $\phi_m$ on $G/H$ is called a symmetric field if its value 
 at $y^\prime \in G/H$ is related to the value at any other point $y \in G/H$
through a local gauge transformation
{ $U(y, g_0) \in H \subset G_{\rm YM}$} and a local Lorentz transformation 
{$\Lambda_{mn}(y, g_0)$}
as
\begin{equation}
\phi_m(y')\ =\ \Lambda_{mn} U  \phi_n(y)U^\dag, 
\end{equation}
 where $g_0$ is an isometry of $G/H$ relating the points $y, y'$ \cite{Forgacs:1979zs}.
It is a natural generalization of a constant field on a flat space to a coset space.
Restricting a higher dimensional theory on symmetric fields on $G/H$ corresponds 
 to focussing on invariant functions under an isometry of $G/H$ up to local symmetry transformations. 
This criterion is based on the expectation
 that the lowest energy field configuration is the most symmetric one,
 and the coset space dimensional reduction retaining only symmetric fields
 is a natural generalization of the ordinary dimensional reduction
 retaining only constant modes on a flat torus. 
Non-constant modes, i.e., non-symmetric fields, 
 correspond to massive fields whose excitation typically costs some amount of energy. 

%\vspace{5mm}

In this paper, 
instead of restricting the higher dimensional Yang-Mills theory on $G/H$ to only the low lying states, 
we will keep all higher Kaluza-Klein modes and investigate their
important roles in restoration of gauge symmetries, which
would be spontaneously broken by  condensation of a symmetric field. 
In particular, we show that some higher excited states become massless under the condensation 
of a tachyonic symmetric field. 

%\vspace{5mm}

Let us consider a field $\phi_m$  satisfying the condition
\begin{equation}
\bar{D}_\alpha \phi_m\ =\ 0, \hspace{1cm} \partial_\alpha \phi_m\ =\ 0. 
   \label{symmetric Higgs def}
\end{equation}
This turns out to be a symmetric field. 
See Appendix \ref{app: symmetric Higgs field} for the proof. 
We call such a field a {\it symmetric Higgs field}. 
The name comes from the fact that
 the field satisfying the above conditions always has a negative mass squared
 (\ref{mass of symmetric Higgs appendix}), as shown in Appendix \ref{app: laplacian and mass formula},
 and develops a vacuum expectation value (vev), which would lead to spontaneous gauge symmetry breaking. 

For the symmetric Higgs fields, the scalar potential of $S_3$ in (\ref{scalar potential original})
becomes simplified as 
\begin{equation}
V(\phi)\ =\ \frac14{\rm Tr}\left( \bar{F}_{mn}+i[\phi_m,\phi_n] \right)^2, 
   \label{scalar potential algebraic}
\end{equation}
where the background field strength (\ref{fieldstrength}) is
\begin{equation}
\bar{F}_{mn}\ =\ f^a{}_{mn}T_a.
\end{equation}
It is nonvanishing only for $ T_a \in \mathfrak{h} \subset \mathfrak{g}_{\rm YM}$,
 where $\mathfrak{h}$ is the Lie algebra of $H$.

Recalling the expression for the covariant derivative (\ref{covariant derivative general}),
 we find that the defining relations (\ref{symmetric Higgs def}) of symmetric Higgs fields imply
\begin{equation}
[T_a,\phi_m]\ =\ if^n{}_{am}\phi_n. 
   \label{condition symmetric Higgs}
\end{equation}
Note that $T_a$ are generators of $G_{\rm YM}$, 
 while $f^n{}_{am}$ are structure constants of the Lie algebra $\mathfrak{g}$ of $G$,
 not those of $\mathfrak{g}_{\rm YM}$. 
Comparing (\ref{condition symmetric Higgs}) with the second equation of (\ref{symmetric coset commutators}), 
 we can see that $\phi_m$ is expressed by the representation $R_t$ of $H$, possibly with a multiplicity.
Thus we can write
 $\phi_m$ of a symmetric Higgs field as 
\begin{equation}
\phi_m(x)\ =\   \varphi_s(x)T_m^s, 
   \label{symmetric Higgs in generator}
\end{equation} 
where $T^s_m$ are generators of $\mathfrak{g}_{\rm YM}$ satisfying
$ 
[T_a,T_m^s]=if^n{}_{am}T_n^s. 
$ 
Note that $T^s_m$ are different generators for different $s$, as we will see in the following sections.
To find the expression for a symmetric Higgs field,
 we decompose the adjoint representation of $G_{\rm YM}$ into irreducible representations of $H$. 
There could exist  representations isomorphic to $R_t$ in the decomposition. 
In the following sections, we will explicitly investigate this in various examples.

%%%%%%%%%%%%%%%%%%%%%%%%%%%%%%%%%%%%%%%%%%%%
%%%%%%%%%%%%%%%%%%%%%%%%%%%%%%%%%%%%%%%%%%%%
%%%%%%%%%%%%%%%%%%%%%%%%%%%%%%%%%%%%%%%%%%%%
%%%%%%%%%%%%%%%%%%%%%%%%%%%%%%%%%%%%%%%%%%%%

\vspace{1cm}

\section{Higgs condensation on $S^2={\rm SU}(2)/{\rm U}(1)$}
\label{condensation  S^2}

\vspace{5mm}

In this section, we consider ${\rm SU}(3)$ Yang-Mills theory compactified on
 the coset space $S^2={\rm SU}(2)/{\rm U}(1)$.  Thus, $G_{\rm YM}={\rm SU}(3)$, $G={\rm SU}(2)$ and $H={\rm U}(1)$.
We choose the generators of $su(2)$ such that the commutation relations are 
\begin{equation}
[t_3,t_\pm]\ =\ \pm t_\pm, \hspace{1cm} [t_+,t_-]\ =\ 2t_3. 
\label{su(2)commutation relation}
\end{equation}
Then, the index $m$ for the tangent space takes $+$ and $-$. 
We denote the generator of $u(1)$ embedded into $su(3)$ by $T$. 
The background gauge field  is only present in the subgroup $H={\rm U}(1)$, and
the scalar potential (\ref{scalar potential algebraic}) becomes 
\begin{eqnarray}
V(\phi) 
&=& -\frac1{8}{\rm Tr}\left( \bar{F}_{+-}+i[\phi_+,\phi_-] \right)^2 \nonumber \\ [2mm] 
&=& \frac1{8}{\rm Tr}\left( 2T-[\phi_+,\phi_-] \right)^2. 
   \label{scalar potential S^2}
\end{eqnarray}
In the following, we will show that different choices of $T$ give us different contents of symmetric Higgs fields with different patterns of their condensation. 

When the coset space is $S^2$, 
the background gauge field $\bar{A}_\alpha$ and the zweibein $e^m_\alpha$ can be explicitly written
as reviewed in Appendix  \ref{app: explicit formulas}.
In fact, $\bar{A}_\alpha$ is given by the monopole configuration on $S^2$ embedded into ${\rm SU}(3)$ gauge group. 
Details on these expressions, in addition to the explicit formula for the covariant derivative $\bar{D}_m\phi_n$, can be found 
in Appendix \ref{app: explicit formulas}. 
We can use these explicit expressions, in particular,
the monopole harmonics \cite{Wu:1976ge} to investigate  the spectrum in the Kaluza-Klein reduction. 
However it will turn out that 
a more abstract formalism \cite{Salam:1981xd} reviewed in Appendix \ref{mode functions appendix} is sufficient for the purpose
since various analytic calculations can be reduced to group-theoretic arguments on the coset space. 
Such an abstract formalism is straightforwardly extended 
to more general coset spaces, such as $\mathbb{CP}^2={\rm SU}(3)/({\rm SU}(2)\times{\rm U(1)})$ which will be discussed in the next section.

%%%%%%%%%%%%%%%%%%%%%%%%%%%%%%%%%%%%%%%%%%%%%%

\vspace{5mm}

\subsection{Embedding of $H=\,$U$(1)$ into $G_{\rm YM}=\,$SU$(3)$: Case 1}
\label{S^2 1}

\vspace{5mm}

Our first choice of the embedding of the $H=\rm U(1)$ generator $T$ in $G_{\rm YM}=\rm SU(3)$ is 
\begin{equation}
T\ =\ \frac12\left[
\begin{array}{ccc}
1 & 0 & 0 \\
0 & -1 & 0 \\
0 & 0 & 0 
\end{array}
\right]. 
   \label{T for S^2 1}
\end{equation}
The background flux $\bar{F}_{+-}=-2iT$ breaks the gauge group ${\rm SU}(3)$ to its Cartan subgroup ${\rm U}(1)\times{\rm U}(1)$. 

Let us now find the symmetric Higgs field satisfying the  relation  (\ref{condition symmetric Higgs}) for 
the U(1) generator $T$. 
We first define $T$-charges $q_{ij}$ of $\phi_{\pm,ij}$ fields by 
\begin{equation}
[T,\phi_{\pm}]_{ij}\ =\ q_{ij}\,\phi_{\pm,ij}, 
   \label{T-charge definition}
\end{equation}
where $i,j =1,2,3$ are indices of $3 \times 3$ matrices, and no summation is taken. 
For the choice (\ref{T for S^2 1}) of $T$, the $T$-charges are given in the matrix notation as  
\begin{equation}
q\ =\ \left[
\begin{array}{ccc}
0 & 1 & \frac12 \\
-1 & 0 & -\frac12 \\
-\frac12 & \frac12 & 0
\end{array}
\right]. 
   \label{charge 1}
\end{equation}
Then, the condition (\ref{condition symmetric Higgs}) for the symmetric Higgs field
and the commutation relation (\ref{su(2)commutation relation}) 
tell us that the $(i,j)=(1,2)$ and $(2,1)$ components of  $\phi_{\pm,ij}$ 
with $T$-charge  $\pm1$ provide us with the symmetric Higgs fields. 
Thus  there is only one symmetric Higgs field (and its complex conjugate)  given by
\begin{equation}
\phi_+(x) \ =\ \left[
\begin{array}{ccc}
0 & \varphi (x)& 0 \\
0 & 0 & 0 \\
0 & 0 & 0
\end{array}
\right], \hspace{1cm}  
\phi_-(x) \ =\ \left[
\begin{array}{ccc}
0 & 0 & 0 \\
\varphi^\dag (x) & 0 & 0 \\
0 & 0 & 0
\end{array}
\right], 
\end{equation}
where we have used $\phi_-=(\phi_+)^\dag$. 

Inserting these expressions into the scalar potential (\ref{scalar potential S^2}), we obtain 
the scalar potential for the symmetric Higgs field $\varphi$
\begin{equation}
V(\phi)\ =\ \frac14\left( 1-|\varphi|^2 \right)^2.
\end{equation}
Thus $\varphi$ will acquire vev at $|\varphi|=1$. 
At the origin $\varphi=0$, as mentioned before, the gauge symmetry $\rm SU(3)$ is broken to
  ${\rm U}(1)\times{\rm U}(1) \subset {\rm SU}(3)$  by the background gauge flux.
When the symmetric Higgs field acquires vev at $|\varphi|=1$, the gauge symmetry 
is expected to be further broken to ${\rm U}(1)$ by the Higgs mechanism.
Thus the expected symmetry breaking pattern is as follows:
\begin{equation}
  G_{\rm YM}={\rm SU}(3) \xrightarrow{{\rm background}\ {\rm flux}} {\rm U}(1) \times {\rm U}(1)
  \xrightarrow{{\rm Higgs} \ {\rm vev}} {\rm U}(1) \ ?
  \label{SBstrucutreS2-1naive}
\end{equation}

%\vspace{5mm}

This is the  usual argument for the gauge symmetry breaking 
in the context of the coset space dimensional reduction in which
only the low lying states are taken into considerations. 
However,
 the conclusion of the gauge symmetry breaking is suspicious
 in view of the higher dimensional gauge theory with the Kaluza-Klein reduction. 
The reason is the following. 
Note that we have vanishing scalar potential $V({|\varphi|=1})=0$ at the 
the global minimum of $V(\phi)$. 
Since the scalar potential originally comes from the terms (\ref{scalar potential original}), 
the vanishing scalar potential implies that the gauge field $A_\alpha$ at the symmetric Higgs vev $|\varphi|=1$
must be  a pure gauge, and 
we must conclude that the full gauge symmetry ${\rm SU}(3)$ is recovered at the symmetric Higgs vacuum, 
instead of being broken to U(1). 

In the rest of this section, in order to show the restoration of the gauge symmetry, we 
will explicitly see that some of the originally massive Kaluza-Klein vector fields become massless at vev $|\varphi|=1$, 
and  eight massless vector fields emerge at the symmetric Higgs vacuum. 
These massless vector fields are the gauge fields due to the general argument by Weinberg \cite{Weinberg:1995mt}. 

%\vspace{5mm}

Mass term of the vector field  $a_\mu$ comes from the term (\ref{mass term vector}), 
and a vector field is massless in the presence of the symmetric Higgs vev if and only if
\begin{equation}
\bar{D}_+a_\mu(x,y) +i[T_+,a_\mu(x,y)]\ =\ 0, \hspace{1cm} 
T_+\ :=\ \left[
\begin{array}{ccc}
0 & 1 & 0 \\
0 & 0 & 0 \\
0 & 0 & 0
\end{array}
\right]
   \label{condition massless 1}
\end{equation}
is satisfied. 
Note that 
$T$, $T_+$ and $T_-:=(T_+)^\dag$ form an $su(2)$ subalgebra of $su(3)$, 
and $a_\mu$ is in the  adjoint representation $\bf 8$ of $su(3)$. 
Thus,
 by the irreducible decomposition of $\bf 8$ of $su(3)$ into $\bf 3\oplus2\oplus2'\oplus1$ of $su(2)$, 
 the condition (\ref{condition massless 1}) can be decomposed into the following four conditions.
First, for the representation $\bf 3$,   we have
\begin{equation}
[1]\ {\rm Massless \ Cond. \ for \ {\bf 3}}  \hspace{10mm}
\bar{D}_+\left[
\begin{array}{c}
a_{\mu,12} \\ [1mm] 
a_{\mu,11}-a_{\mu,22} \\ [1mm]
a_{\mu,21}
\end{array}
\right]\ =\ -i\left[
\begin{array}{c}
a_{\mu,22}-a_{\mu,11} \\ [1mm] 
2a_{\mu,21} \\ [1mm] 
0
\end{array}
\right].
   \label{condition massless vector spin 1}
\end{equation}
For the representations $\bf 2$ and $\bf 2'$, we have
\begin{equation}
[2]\  {\rm Massless \ Cond. \ for} \ {\bf 2}  \hspace{10mm}
\bar{D}_+\left[
\begin{array}{c}
a_{\mu,13} \\ [1mm]
a_{\mu,23}
\end{array}
\right]\ =\ -i\left[
\begin{array}{c}
a_{\mu,23} \\ [1mm]
0
\end{array}
\right]
   \label{condition massless vector spin 1/2}
\end{equation}
 with the condition for their conjugate components $a_{\mu,31}, a_{\mu,32}$ which is equivalent to 
\begin{equation}
[2^\prime]\  {\rm Massless \ Cond. \ for} \ {\bf 2'}   \hspace{10mm}
\bar{D}_-\left[
\begin{array}{c}
a_{\mu,13} \\ [1mm]
a_{\mu,23}
\end{array}
\right]\ =\ -i\left[
\begin{array}{c}
0 \\ [1mm] 
a_{\mu,13}
\end{array}
\right], 
   \label{condition massless vector spin 1/2-2}
\end{equation}
and finally, 
\begin{equation}
[3]\  {\rm Massless \ Cond. \ for \  {\bf 1}}  \hspace{10mm}
\bar{D}_+a_{\mu,33}\ =\ 0
   \label{condition massless vector spin 0}
\end{equation}
for the {singlet} representation $\bf 1$. 
A vector field satisfying one of these conditions 
become massless at the global minimum of the Higgs potential $V(\varphi)$
at $|\varphi|=1$.
These are a set of first-order partial differential equations which can be written explicitly by using the formulas in Appendix 
\ref{app: explicit formulas}, and 
the number of massless vector fields can be found by solving the above differential equations. 
In the following, instead of solving them explicitly, 
we solve these conditions by reducing  to a group-theoretic problem. 

%\vspace{5mm}

For this purpose, we need to understand the action of the covariant derivative $\bar{D}_\pm$ on $a_\mu$ \cite{Salam:1981xd}. 
The action can be simplified by choosing a suitable complete set of functions on $S^2$
which can be used to expand $a_\mu(x,y)$. 
Generally speaking, as explained in Appendix \ref{mode functions appendix}, 
due to the Peter-Weyl theorem, 
a complete set of functions on a group manifold $G$ is given by the representation matrices
$\rho^R(g)_{IJ}$ for all the representation $R$ {of $G$} and their components $I,J=1, \cdots , \dim R$.
Then a complete set of functions on $G/H$ is obtained by imposing particular transformation laws under $H$,
 corresponding to the $T$-charge of functions on $G/H$.
Collecting all the representations of $H$, the complete set on $G$ is recovered. 

In the case of $S^2={\rm SU}(2)/{\rm U}(1)$, 
 a complete set on $S^2$, collecting all the charges of $H={\rm U}(1)$, is given by
\begin{equation}
f^{j}_{mm'}(y) \hspace{5mm}  {\rm where} \ \   j\ =\ 0,\ \frac12,\ 1,\ \frac32,\ \cdots, \hspace{5mm} -j\ \le\ m,m'\,\le\ j. 
\end{equation}
Each $j$ corresponds to the spin $j$ representation of ${\rm SU}(2)$.  
As explained in  Appendix \ref{mode functions appendix}, 
the function $f^j_{mm'}$ has  $T$-charge $m$ of $H={\rm U}(1)$. 
Thus, $m=0$ gives the  usual spherical harmonics, while $m \neq 0$ modes
are the monopole spherical harmonics with $T$-charge $m$, which are relevant in the monopole background. 

A field $\chi(y)$ on $S^2$ with the $T$-charge $q$ is then expanded 
in terms of $f^j_{qm'}(y)$ as 
\begin{equation}
\chi(y)\ =\ \sum_j\sum_{m'=-j}^jc^j_{m'}f^j_{qm'}(y), 
\label{S2-Tcharge-q}
\end{equation}
where the sum of $j$ is taken over all values of the spin $j$ whose magnetic quantum number $m'$ can take $q$. 
Explicitly, $j$ in the sum must satisfy 
\begin{equation}
-j\ \le\ q\ \le\ j, \hspace{1cm} j-q\ \in\ \mathbb{Z}. 
\end{equation}
From this expansion, we obtain $2j+1$ complex-valued fields, labeled by $m'$, with the $T$-charge $q$ from each $j$. 

In order to discuss the  massless condition (\ref{condition massless 1}) for vector fields, 
it is sufficient to know the action of $\bar{D}_+$ on the mode functions $f^j_{mm'}(y)$. 
From (\ref{covariant derivative algebraic}) in Appendix \ref{app: laplacian and mass formula}, 
this action turns out to be given by
\begin{equation}
\bar{D}_+f^j_{mm'}(y)\ =\ -i\sum_{n=-j}^j\left( T^{(j)}_+ \right)_{mn}f^j_{nm'}(y), 
   \label{covariant derivative generator S^2}
\end{equation}
where $T^{(j)}_+$ is the spin-$j$ representation of $t_+$. 
Note that it is valid irrespective of the value of the symmetric Higgs field. 
Thus,  the condition (\ref{condition massless 1}) for massless vector fields is reduced to 
algebraic relations of the coefficients $c^j_{\mu,i_1i_2,m'}$ in the mode expansion 
\begin{equation}
a_{\mu,i_1 i_2}(y)\ =\ \sum_j\sum_{m'=-j}^j c^j_{\mu,i_1i_2,m'}f^j_{q(i_1,i_2),m'}(y)
\end{equation}
between the first and the second terms in (\ref{condition massless 1}). 
Here $q(i_1,i_2)$ is the $T$-charge of $(i_1,i_2)$-component of $a_\mu$.
The first term in  (\ref{condition massless 1}) 
 is a multiplication of $T^{(j)}_+$ on the complete set $f^j_{mm'}(y)$ due to the covariant derivative $\bar{D}_+$
while the second term 
is the adjoint action of $T_+$ due to the symmetric Higgs vev.
If we can choose the expansion coefficients of $a_\mu$ such that 
these two actions have the same effect, then we obtain a massless vector field. 

%\vspace{5mm}

Let us check whether this condition can be satisfied. 
First,
 we consider (\ref{condition massless vector spin 1/2}) of the massless condition
 for representation $\bf 2$, i.e., {$q=\pm1/2$}.
Thus, the representations of $\rm SU(2)$ are restricted to be $j=k+1/2$ for non-negative integers $k$. 
This can be written as 
\begin{eqnarray}
\sum_{k=0}^\infty\sum_{m'=-k-\frac12}^{k+\frac12}c^{k}_{m^\prime}\left( T^{(k+\frac12)}_+ \right)_{\frac12,n}f^{k+\frac12}_{nm'}(y) &=& \sum_{k=0}^\infty\sum_{m'=-k-\frac12}^{k+\frac12}\tilde{c}^k_{m'}f^{k+\frac12}_{-\frac12,m'}(y)
\end{eqnarray}
and
\begin{eqnarray}
\sum_{k=0}^\infty\sum_{m'=-k-\frac12}^{k+\frac12}\tilde{c}^k_{m'}\left( T^{(k+\frac12)}_+ \right)_{-\frac12,n}f^{k+\frac12}_{nm'}(y) &=& 0, 
\end{eqnarray}
{where we renamed the coefficients $c^j_{\mu,13, m^\prime}$ and $c^j_{\mu,23,m'}$ as $c^{k}_{m^\prime}$ and $\tilde{c}^k_{m'}$, respectively. }
These equations are satisfied if and only if $c^0_{m'}=\tilde{c}^0_{m'}$ are the only non-zero coefficients. 
Note that in our normalization and notation 
\begin{equation}
\left( T^{(\frac12)}_+ \right)_{\frac12,-\frac12}\ = (\sigma_+)_{12} = \ 1 \, ,
\end{equation}
where $\sigma_+$ is the Pauli matrix.
The same coefficients also solve (\ref{condition massless vector spin 1/2-2}). 
Consequently, we have shown that 
\begin{equation}
a_{\mu,13}(x,y)\ =\ \sum_{m'=\pm\frac12}c^0_{m'}(x)f^{\frac12}_{\frac12,m'}(y), \hspace{1cm} a_{\mu,23}(x,y)\ =\ \sum_{m'=\pm\frac12}c^0_{m'}(x)f^{\frac12}_{-\frac12,m'}(y)
\end{equation}
 and their conjugates give us four massless vector fields. 

%\vspace{5mm}

Next, we consider (\ref{condition massless vector spin 1}) 
for the massless condition for representation $\bf 3$, i.e., {$q=0,\pm1$}.  
Thus, the representations of $\rm SU(2)$ are restricted to be $j=l$ for non-negative integers $l$. 
The range of $l$ depends on the $T$-charges. 
The linear combinations
\begin{equation}
-a_{\mu,12}, \hspace{1cm} \frac1{\sqrt{2}}(a_{\mu,11}-a_{\mu,22}), \hspace{1cm} a_{\mu,21} 
\end{equation}
as the independent fields are convenient for our purpose. 
Then, the condition (\ref{condition massless vector spin 1}) can be written as 
\begin{eqnarray}
\sum_{l=1}^\infty \sum_{m'=-l}^lc^l_{m'}\left( T^{(l)}_+ \right)_{1,n}f^l_{nm'}(y) &=& \sqrt{2}\sum_{l=0}^\infty \sum_{m'=-l}^l \widetilde{c}^l_{m'}f^l_{0,m'}(y), \\
\sum_{l=0}^\infty \sum_{m'=-l}^l\widetilde{c}^l_{m'}\left( T^{(l)}_+ \right)_{0,n}f^l_{nm'}(y) &=& \sqrt{2}\sum_{l=1}^\infty \sum_{m'=-l}^l \widehat{c}^l_{m'}f^l_{-1,m'}(y), \\
\sum_{l=1}^\infty \sum_{m'=-l}^l\widehat{c}^l_{m'}\left( T^{(l)}_+ \right)_{-1,n}f^l_{nm'}(y) &=& 0. 
\end{eqnarray}
These equations are satisfied if and only if $c^1_{m'}=\widetilde{c}^1_{m'}=\widehat{c}^1_{m'}$ are the only non-zero coefficients. 
Then, the following three combinations
\begin{eqnarray}
-a_{\mu,12} &=& { \sum_{m'=-1,0,1} c^1_{m'} f^1_{1,m'} (y),  } \\ 
\frac1{\sqrt{2}}(a_{\mu,11}-a_{\mu,22}) &=& \sum_{m'=-1,0,1} c^1_{m'} f^1_{0,m'} (y), \\
a_{\mu,21} &=& \sum_{m'=-1,0,1} c^1_{m'}f^1_{-1,m'} (y),   
\end{eqnarray}
with the condition $a_\mu^\dag=a_\mu$, give us 3 massless vector fields. 
In fact, this can be easily anticipated by rewriting (\ref{condition massless vector spin 1}) as 
\begin{equation}  
\bar{D}_+\left[
\begin{array}{c}
-a_{\mu,12} \\ [1mm] 
\frac1{\sqrt{2}}(a_{\mu,11}-a_{\mu,22}) \\ [1mm]
a_{\mu,21}
\end{array}
\right]\ =\ -iT^{(1)}_+\left[
\begin{array}{c}
-a_{\mu,12} \\ [1mm] 
\frac1{\sqrt{2}}(a_{\mu,11}-a_{\mu,22}) \\ [1mm]
a_{\mu,21}
\end{array}
\right]. 
\end{equation}
Namely, these three components form the triplet of the $su(2)$ subalgebra, as mentioned before. 
Note that $a_{\mu,11}-a_{\mu,22}$ has also a contribution from $j=0$ which was massless before the Higgs condensation. 
This becomes massive due to the Higgs mechanism. 

%\vspace{5mm}

Finally, let us consider the condition (\ref{condition massless vector spin 0}) 
for the massless condition for representation $\bf 1$, i.e., $q=0$.  
Thus, the representation $f^j_{m,m^\prime}$ of $SU(2)$ is restricted to be $j=l$ for non-negative integers $l$. 
The condition simply means that $a_{\mu,33}$ is independent of $y$, resulting in one massless vector field. 
This is nothing but the ${\rm U}(1)$ gauge field which is unbroken after the Higgs condensation. 

%\vspace{5mm}

In total, we have found {\bf eight massless vector fields}
 which should correspond to the ${\rm SU}(3)$ gauge field which is expected to appear at the symmetric Higgs vacuum. 
 Therefore we conclude that, contrary to the expectation in (\ref{SBstrucutreS2-1naive}) 
 within the analysis of the low lying states, 
  the symmetry breaking-restoration pattern is given by
 \begin{equation}
  G_{\rm YM}={\rm SU}(3) \xrightarrow{{\rm background}\ {\rm flux}} {\rm U}(1) \times {\rm U}(1)
  \xrightarrow{{\rm Higgs} \ {\rm vev}} {\rm SU}(3) \ 
\end{equation}
if all the Kaluza-Klein modes are taken into considerations. 
It should be noted that seven massless vector fields out of eight ones come from the Kaluza-Klein modes which were massive before the Higgs condensation. 
Indeed, all the massless vector fields before the Higgs condensation come from $j=0$ mode in the expansions, while seven massless fields after the Higgs condensation come from the $j=\frac12$ and $j=1$ modes. 
This phenomenon happens because we keep all the Kaluza-Klein modes in the model, in contrast to the 
simplest analysis of the coset space dimensional reduction in which only the low lying modes are taken into account. 
Usually, the Kaluza-Klein modes are considered to be so heavy that they are not considered in discussing the dynamics of light fields. 
However, since our model has only a single mass scale given by the radius of the coset space, 
the potential height and the Higgs vev are also of the same Kaluza-Klein mass scale. 
This enables some of the massive Kaluza-Klein modes to become massless. 

The investigation of the massless vector fields performed above is possible 
because the vev of $\varphi$ is exactly the right value so that the vev of $\phi_+$ is equal to the generator $T_+$, 
including the overall normalization. 
We will observe in the following that this coincidence persists for the other models discussed in this paper. 
It is very interesting to clarify whether this is the general feature of the Kaluza-Klein reduction on coset spaces. 
If this is the case, the masses of the fields at the symmetric Higgs vacuum would be possibly given in terms of group-theoretic quantities, 
as  the mass formulas in Appendix \ref{app: laplacian and mass formula}
 are valid before the symmetric Higgs condensation.

%%%%%%%%%%%%%%%%%%%%%%%%%%%%%%%%%%%%%
%%%%%%%%%%%%%%%%%%%%%%%%%%%%%%%%%%%%%
%%%%%%%%%%%%%%%%%%%%%%%%%%%%%%%%%%%%%

\vspace{5mm}

\subsection{Embedding of $H=\,$U$(1)$ into $G_{\rm YM}=\,$SU$(3)$: Case 2}
\label{S^2 2}

\vspace{5mm}

Our next choice of an embedding of U(1) charge $T$ into $G_{\rm YM}=$SU(3) is 
\begin{equation}
T\ =\ \frac13\left[
\begin{array}{ccc}
1 & & \\
 & 1 & \\
 & & -2
\end{array}
\right]. 
\end{equation}
The corresponding background flux $\bar{F}_{+-}=-2iT$ breaks the gauge group $G_{\rm YM}={\rm SU}(3)$ 
into ${\rm SU}(2)\times {\rm U}(1)$. 

The $T$-charges for the components of $\phi_+$ defined in (\ref{T-charge definition}) are given by
\begin{equation}
q\ =\ \left[
\begin{array}{ccc}
0 & 0 & 1 \\
0 & 0 & 1 \\
-1 & -1 & 0
\end{array}
\right]. 
\end{equation}
Recall that the components of $\phi_+$ with the $T$-charge $+1$ become the symmetric Higgs fields, 
and there are two such components. 
Therefore, the symmetric Higgs fields are given by
\begin{equation}
\phi_+\ =\ \left[
\begin{array}{ccc}
0 & 0 & \varphi_1 \\
0 & 0 & \varphi_2 \\
0 & 0 & 0
\end{array}
\right], \hspace{1cm} 
\phi_-\ =\ \left[
\begin{array}{ccc}
0 & 0 & 0 \\
0 & 0 & 0 \\
\varphi_1^\dag & \varphi_2^\dag & 0 
\end{array}
\right]. 
\label{Higgs phipm}
\end{equation}
The components $\varphi_1$ and $\varphi_2$ form a doublet of the unbroken ${\rm SU}(2)$ gauge group. 

The scalar potential for this symmetric Higgs doublet can be obtained by inserting the above expressions into (\ref{scalar potential S^2}). 
We obtain 
\begin{equation}
V(\phi)\ =\ \frac14\left( 1-|\varphi|^2 \right)^2+\frac1{12}, 
   \label{scalar potential 2}
\end{equation}
where $|\varphi|^2:=|\varphi_1|^2+|\varphi_2|^2$. 
At the minimum of the potential, they acquire the vev 
\begin{equation}
\varphi_1\ =\ 1, \hspace{1cm} \varphi_2\ =\ 0, 
\end{equation}
up to a global SU(2) gauge transformation. 
This would break the gauge group ${\rm SU}(2)\times{\rm U}(1)$ preserved by the background flux to ${\rm U}(1)$. 
Manton \cite{Manton:1979kb} applied 
this mechanism of the gauge symmetry breaking to realization of the Weinberg-Salam model
based on the six-dimensional Yang-Mills theory. The expected symmetry breaking pattern 
within the low lying states would be as follows:
\begin{equation}
  G_{\rm YM}={\rm SU}(3) \xrightarrow{{\rm background}\ {\rm flux}} {\rm SU}(2) \times {\rm U}(1)
  \xrightarrow{{\rm Higgs} \ {\rm vev}} {\rm U}(1) \ ?
  \label{SBstrucutreS2-2naive}
\end{equation}

In order to perform the calculation of $V(\phi)$ while keeping the ${\rm SU}(2)\times{\rm U}(1)$ gauge invariance, 
it is convenient to introduce $3 \times 3$ matrices ${\cal T}_s$ ($s=1,2$) and write 
 $\phi_+$ in (\ref{Higgs phipm}) as 
\begin{equation}
\phi_+\ =\sum_{s=1,2} \varphi_s {\cal T}_s. \hspace{1cm} 
\end{equation}
The matrices ${\cal T}_s$ defined by this relation satisfy 
\begin{equation}
[{\cal T}_s,{\cal T}_t^\dag]\ =\ \sigma^-_{ts}\left[
\begin{array}{ccc}
0 & 1 & 0 \\
0 & 0 & 0 \\
0 & 0 & 0
\end{array}
\right]+\sigma^+_{ts}\left[
\begin{array}{ccc}
0 & 0 & 0 \\
1 & 0 & 0 \\
0 & 0 & 0 
\end{array}
\right]+\sigma^3_{ts}\left[
\begin{array}{ccc}
\frac12 & 0 & 0 \\
0 & -\frac12 & 0 \\
0 & 0 & 0 
\end{array}
\right]+\frac32\delta_{ts}T, 
\end{equation}
where $\sigma^+_{ts}$ etc. are $(t,s)$-components of the Pauli matrices.

%\vspace{5mm}

Let us count the number of massless vector fields at the symmetric Higgs vacuum. 
We notice that the vev of $\phi_+$ is 
\begin{equation}
\phi_+\ =\ \left[
\begin{array}{ccc}
0 & 0 & 1 \\
0 & 0 & 0 \\
0 & 0 & 0
\end{array}
\right]
\end{equation}
which is the spin-$\frac12$ representation of $t_+$ embedded into an $su(2)$ subalgebra of $su(3)$ different from the one in the previous section. 
The massless conditions in this case can be obtained from the previous ones by simply exchanging $2$ and $3$ in the matrix indices. 
For example, we have 
\begin{equation}
\bar{D}_-\left[
\begin{array}{c}
a_{\mu,12} \\ [1mm]
a_{\mu,32}
\end{array}
\right]\ =\ -i\left[
\begin{array}{c}
0 \\ [1mm] 
a_{\mu,12}
\end{array}
\right].
   \label{condition massless vector spin 1/2 2nd}
\end{equation}
In this case, however, the $T$-charges of the components are different. 
The $T$-charges for $a_\mu$ are 
\begin{equation}
q\ =\ \left[
\begin{array}{ccc}
0 & 0 & 1 \\
0 & 0 & 1 \\
-1 & -1 & 0
\end{array}
\right]. 
\end{equation}
Therefore, $a_{\mu,12}$ has the $T$-charge 0, while $a_{\mu,32}$ has the $T$-charge $-1$. 
Their mode expansions are given as 
\begin{eqnarray}
a_{\mu,12} &=& \sum_{j=0}^\infty \sum_{m'=-j}^j c^j_{m'}f^j_{0,m'}, \\
a_{\mu,32} &=& \sum_{j=1}^\infty \sum_{m'=-j}^j \tilde{c}^j_{m'}f^j_{-1,m'}. 
\end{eqnarray}
We find that $\bar{D}_-f^j_{0,m'}$ vanishes only if $j=0$. 
Since $\bar{D}_-a_{\mu,32}$ does not have a contribution from the spin-0 representation, we conclude that the condition (\ref{condition massless vector spin 1/2 2nd}) does not have a solution. 

We also have 
\begin{equation}
\bar{D}_+\left[
\begin{array}{c}
-a_{\mu,13} \\ [1mm] 
\frac1{\sqrt{2}}(a_{\mu,11}-a_{\mu,33}) \\ [1mm]
a_{\mu,31}
\end{array}
\right]\ =\ -iT^{(1)}_+\left[
\begin{array}{c}
-a_{\mu,13} \\ [1mm] 
\frac1{\sqrt{2}}(a_{\mu,11}-a_{\mu,33}) \\ [1mm]
a_{\mu,31}
\end{array}
\right]. 
\end{equation}
The $T$-charge assignment for $a_\mu$ turns out to be appropriate so that we can find the following solution 
\begin{eqnarray}
-a_{\mu,13}(x,y) &=& {\sum_{m^\prime =-1}^1 c^1_{m'} f^1_{1,m^\prime}(y), } \\  %c^1_{m'}
\frac1{\sqrt{2}}(a_{\mu,11}(x,y)-a_{\mu,33}(x,y)) &=& {\sum_{m'=-1}^1c^1_{m'}f^1_{0,m'} (y),  }\\ %c^1_{m'}
 a_{\mu,13}(x,y) &=& { \sum_{m'=-1}^1c^1_{m'}f^1_{-1,m'} (y).  }
\end{eqnarray}
They give us three massless vector fields. 

The last condition 
\begin{equation}
\bar{D}_+a_{\mu,22}\ =\ 0
\end{equation}
gives us one massless vector field. 

In total, we have found {\bf four massless vector fields}. 
Three of them were massive before the Higgs condensation. 
Since scalar potential (\ref{scalar potential 2})
 does not vanish at the symmetric Higgs vacuum $|\varphi|=1$, 
 there remains a non-trivial flux after the Higgs condensation
 which prevents the full ${\rm SU}(3)$ gauge symmetry from recovering. 
Probably, the gauge group at the symmetric Higgs vacuum would be ${\rm SU}(2)\times{\rm U}(1)$, where the ${\rm SU}(2)$ part is not the one preserved by $\bar{F}_{+-}$ but an ``emergent'' one. 
Therefore we may conclude that, contrary to the expectation in (\ref{SBstrucutreS2-2naive}) 
 within the analysis of the low lying states, 
  the symmetry breaking-restoration pattern is given by
 \begin{equation}
  G_{\rm YM}={\rm SU}(3) \xrightarrow{{\rm background}\ {\rm flux}} {\rm SU}(2) \times {\rm U}(1)
  \xrightarrow{{\rm Higgs} \ {\rm vev}} {\rm SU}(2) \times {\rm U}(1) \, . 
\end{equation}
For confirmation of this pattern, a more detailed analysis will be necessary.

%\vspace{5mm}

It is natural to ask whether the symmetric Higgs vacuum $|\varphi|=1$ is stable or not. 
In the previous section, the stability is obvious since the vacuum attains the global minimum of the scalar potential 
in every direction of the field space. 
For the case in this section, it is possible that there still exists a Higgs field at the symmetric Higgs vacuum, and a further condensation would occur. 

At least, we can show that $|\varphi|=1$ is a classical solution of the full theory including all Kaluza-Klein modes. 
In other words, we claim that the symmetric Higgs vacuum discussed in this section has the same relevance as the trivial solution before the symmetric Higgs condensation which has been discussed in the literature \cite{Schellekens:1984dm}. 
To show this, we need to confirm that the symmetric Higgs vev does not act as a source for other scalar fields coming from the Kaluza-Klein expansion of $\phi_\pm$. 
If there would exist terms in the scalar potential of the form 
\begin{equation}
{\rm Tr}(\bar{F}\varphi\Phi), \hspace{1cm} {\rm Tr}(\varphi^3\Phi), 
   \label{source terms}
\end{equation}
where $\Phi$ indicates scalar fields other that the symmetric Higgs fields, then the vev of $\varphi$ would give a source term of $\Phi$, so that $\Phi=0$ is not the classical solution. 
As mentioned above, there could exist terms with $\Phi^2$ which would indicate the presence of other Higgs fields. 
Since this allows $\Phi=0$ to be a classical solution, we ignore them in the following. 

Recall that
 a symmetric Higgs field $\varphi$ is a constant mode on $S^2$ and is singlet for the $H$ transformation,
 which are implied by the conditions (\ref{symmetric Higgs def}).
Then, $\Phi$ is a non-constant mode on $S^2$ or $H$-non-singlet. 
This implies that the terms of the second kind in (\ref{source terms}) is absent. 
Indeed, if $\Phi$ is a non-constant mode, then the integration of ${\rm Tr}(\varphi^3\Phi)$ over $S^2$, performed in the Kaluza-Klein reduction, vanishes due to the orthogonality condition for the mode functions $f^j_{mm'}$. 
On the other hand, if $\Phi$ is $H$-non-singlet, then ${\rm Tr}(\varphi^3\Phi)$ simply vanishes since the scalar potential is $H$-singlet. 
The terms of the first kind in (\ref{source terms}) are also prohibited since the background flux $\bar{F}_{+-}$ is also constant on $S^2$ and $H$-singlet. 
The latter is valid since $\bar{F}_{+-}$ is invariant under the ${\rm U}(1)$ gauge transformation, and is also invariant under the local Lorentz transformation.

%%%%%%%%%%%%%%%%%%%%%%%%%%%%%%%%%%%%%%%%%%%%%%%%
%%%%%%%%%%%%%%%%%%%%%%%%%%%%%%%%%%%%%%%%%%%%%%%%
%%%%%%%%%%%%%%%%%%%%%%%%%%%%%%%%%%%%%%%%%%%%%%%%
\vspace{5mm}

\subsection{Embedding of $H=\,$U$(1)$ into $G_{\rm YM}=\,$SU$(3)$: Case 3}

\vspace{5mm}

Our last choice for $T$ is 
\begin{equation}
T\ =\ \left[
\begin{array}{ccc}
1 & & \\
 & 0 & \\
 & & -1
\end{array}
\right]. 
\end{equation}
The background flux $\bar{F}_{+-}=-2iT$ breaks the gauge group ${\rm SU}(3)$ to ${\rm U}(1)\times{\rm U}(1)$. 
The $T$-charge of the components of $\phi_+$ is then 
\begin{equation}
q\ =\ \left[
\begin{array}{ccc}
0 & 1 & 2 \\
-1 & 0 & 1 \\
-2 & -1 & 0 
\end{array}
\right]. 
   \label{T-charge 3}
\end{equation}
Therefore, there are two symmetric Higgs fields which are given by
\begin{equation}
\phi_+\ =\ \left[
\begin{array}{ccc}
0 & \varphi_1 & 0 \\
0 & 0 & \varphi_2 \\
0 & 0 & 0
\end{array}
\right], \hspace{1cm} 
\phi_-\ =\ \left[
\begin{array}{ccc}
0 & 0 & 0 \\
\varphi_1^\dag & 0 & 0 \\
0 & \varphi_2^\dag & 0
\end{array}
\right]. 
\end{equation}
This is our first example where two independent symmetric Higgs fields appear. 
In the previous section, we also have two symmetric Higgs fields, but they form a doublet of ${\rm SU}(2)$ gauge group. 

The scalar potential for this case becomes 
\begin{equation}
V(\phi)\ = \frac1{8}\left( |\varphi_1|^2-2 \right)^2+\frac1{8}\left( |\varphi_2|^2-|\varphi_1|^2 \right)^2+\frac1{8}\left( |\varphi_2|^2-2 \right)^2. 
\end{equation}
The Higgs vacuum corresponds to $\varphi_1=\varphi_2=\sqrt{2}$ up to ${\rm U}(1)\times{\rm U}(1)$ transformation. 
This attains the global minimum of the scalar potential which implies that the original ${\rm SU}(3)$ gauge symmetry should be recovered. 
Then at the minimum, the symmetric Higgs field $\phi_+$ has the vev given by
\begin{equation}
\phi_+\ = T^{(1)}_+:=
\left[
\begin{array}{ccc}
0 & \sqrt{2} & 0 \\
0 & 0 & \sqrt{2} \\
0 & 0 & 0
\end{array}
\right]. 
\label{T1def}
\end{equation} 
This is the spin-1 representation $T^{(1)}_+$ of $t_+$ embedded into $su(3)$. 
Then the counting in this case is also reduced to a group-theoretic calculation. 

The massless condition (\ref{condition massless 1}),
 in which $T_+$ is replaced with $T^{(1)}_+$, can be written as 
\begin{eqnarray}
& & \bar{D}_+\left[
\begin{array}{ccc}
a_{\mu,11} & a_{\mu,12} & a_{\mu,13} \\
a_{\mu,21} & a_{\mu,22} & a_{\mu,23} \\
a_{\mu,31} & a_{\mu,32} & a_{\mu,33} 
\end{array}
\right] \nonumber \\ [2mm]
&=& -i\left[
\begin{array}{ccc}
\sqrt{2}a_{\mu,21} & \sqrt{2}(a_{\mu,22}-a_{\mu,11}) & \sqrt{2}(a_{\mu,23}-a_{\mu,12}) \\
\sqrt{2}a_{\mu,31} & \sqrt{2}(a_{\mu,32}-a_{\mu,21}) & \sqrt{2}(a_{\mu,33}-a_{\mu,22}) \\
0 & -\sqrt{2}a_{\mu,31} & -\sqrt{2}a_{\mu,32}
\end{array}
\right].
\end{eqnarray}
This can be rearranged into two sets of equations. 
One is 
\begin{eqnarray}
\bar{D}_+\left[
\begin{array}{c}
\sqrt{2}a_{\mu,13} \\
a_{\mu,23}-a_{\mu,12} \\
\frac1{\sqrt3}(a_{\mu,11}-2a_{\mu,22}+a_{\mu,33}) \\
a_{\mu,21}-a_{\mu,32} \\
\sqrt2 a_{\mu,31}
\end{array}
\right] 
&=& -iT^{(2)}_+\left[
\begin{array}{c}
\sqrt{2}a_{\mu,13} \\
a_{\mu,23}-a_{\mu,12} \\
\frac1{\sqrt3}(a_{\mu,11}-2a_{\mu,22}+a_{\mu,33}) \\
a_{\mu,21}-a_{\mu,32} \\
\sqrt2 a_{\mu,31}
\end{array}
\right], \nonumber \\
   \label{condition massless vector spin 2}
\end{eqnarray}
where 
\begin{equation}
T^{(2)}_+\ := \left[
\begin{array}{ccccc}
0 & 2 & 0 & 0 & 0 \\
0 & 0 & \sqrt{6} & 0 & 0 \\
0 & 0 & 0 & \sqrt{6} & 0 \\
0 & 0 & 0 & 0 & 2 \\
0 & 0 & 0 & 0 & 0
\end{array}
\right]
\end{equation}
is the spin-2 representation of $t_+$. 
The other is 
\begin{eqnarray}
\bar{D}_+\left[
\begin{array}{c}
a_{\mu,23}+a_{\mu,12} \\
a_{\mu,33}-a_{\mu,11} \\
-a_{\mu,21}-a_{\mu,32}
\end{array}
\right] 
&=& -iT^{(1)}_+\left[
\begin{array}{c}
a_{\mu,23}+a_{\mu,12} \\
a_{\mu,33}-a_{\mu,11} \\
-a_{\mu,21}-a_{\mu,32}
\end{array}
\right],
   \label{condition massless vector spin 1 3rd}
\end{eqnarray}
 where $T^{(1)}_+$ is the spin-1 representation of $t_+$ given in Eq.(\ref{T1def}).
These two sets of equations correspond to
 the irreducible decomposition of $\bf 8$ of $su(3)$ into $\bf 5\oplus 3$ of an $su(2)$ subgroup.

Recall that the $T$-charges for $a_\mu$ are given also as (\ref{T-charge 3}). 
Each linear combination in the above equations consists of components with the same $T$-charge,
 as it should be.
We find that
 the spin-2 representation in the mode expansion gives
 the solution for the condition (\ref{condition massless vector spin 2}),
 and the spin-1 representation gives
 the solution for the condition (\ref{condition massless vector spin 1 3rd}). 
They give {\bf eight massless vector fields} at the symmetric Higgs vacuum, 
 and as expected, the symmetry pattern is given by
 \begin{equation}
  G_{\rm YM}={\rm SU}(3) \xrightarrow{{\rm background}\ {\rm flux}} {\rm U}(1) \times {\rm U}(1)
  \xrightarrow{{\rm Higgs} \ {\rm vev}} {\rm SU}(3) \, .
\end{equation}

%%%%%%%%%%%%%%%%%%%%%%%%%%%%%%%%%%%%%
%%%%%%%%%%%%%%%%%%%%%%%%%%%%%%%%%%%%%
%%%%%%%%%%%%%%%%%%%%%%%%%%%%%%%%%%%%%

\vspace{1cm}

\section{Higgs condensation on $\mathbb{CP}^2={\rm SU}(3)/({\rm SU}(2)\times{\rm U}(1))$}
\label{condensation CP^2}

\vspace{5mm}

In this section, we consider Yang-Mills theories on $\mathbb{R}^4\times\mathbb{CP}^2$ with the gauge group SU(4).
Since $\mathbb{CP}^2$ can be represented as a coset space $G/H={\rm SU}(3)/({\rm SU}(2)\times{\rm U}(1))$, 
we can apply to this case techniques similar to those  in the previous section. 

The choice of the $H={\rm SU}(2)\times{\rm U}(1)$ subgroup of $G={\rm SU}(3)$
 is specified by their generators given as 
\begin{equation}
t_a\ :=\ \frac12\left[
\begin{array}{cc}
\hspace{3mm}\sigma_a &  
\begin{array}{c}
0 \\
0
\end{array} \\
\begin{array}{cc}
0 & 0
\end{array}
\hspace{-3mm} & 0 
\end{array}
\right],
\hspace{1cm}
t_4\ :=\ \frac13\left[
\begin{array}{ccc}
1 & 0 & 0 \\
0 & 1 & 0 \\
0 & 0 & -2
\end{array}
\right].
   \label{definition coset CP^2}
\end{equation}
 where $a=1,2,3$.
We choose the other generators of $su(3)$ as 
\begin{equation}
t_z\ :=\ \left[
\begin{array}{ccc}
0 & 0 & 1 \\
0 & 0 & 0 \\
0 & 0 & 0 \\
\end{array}
\right], \hspace{1cm} 
t_w\ :=\ \left[
\begin{array}{ccc}
0 & 0 & 0 \\
0 & 0 & 1 \\
0 & 0 & 0 
\end{array}
\right], \hspace{1cm} t_{\bar{s}}\ :=\ (t_s)^\dag, 
 \label{definition coset CP^2-2}
\end{equation}
 where $s=z,w$.
According to this choice, the tangent space index $m$ takes $z,w,\bar{z}$ and $\bar{w}$.

In the following, we always embed $su(2)$ part, $t_1, t_2, t_3$, into  $\mathfrak{g}_{\rm YM}=su(4)$ as 
\begin{equation}
T_a\ :=\ \left[
\begin{array}{cccc}
 & & & 0 \\
 & t_a & & 0 \\
 & & & 0 \\
0 & 0 & 0 & 0
\end{array}
\right]. 
\label{SU(2)embedding}
\end{equation}
In the following,
 we will study two cases of  embedding $T_4$ charge of the $u(1)$ generator, $t_4$, into $su(4)$. 

The scalar potential (\ref{scalar potential algebraic}) becomes 
\begin{eqnarray}
V(\phi) 
&=& \frac1{8}{\rm Tr}\left| \bar{F}_{z\bar{z}}+i[\phi_z,\phi_{\bar{z}}] \right|^2+\frac1{8}{\rm Tr}\left| \bar{F}_{w\bar{w}}+i[\phi_w,\phi_{\bar{w}}] \right|^2 \nonumber \\ [2mm] 
& & +\frac1{4}{\rm Tr}\left| \bar{F}_{zw}+i[\phi_z,\phi_w] \right|^2+\frac1{4}{\rm Tr}\left| \bar{F}_{z\bar{w}}+i[\phi_z,\phi_{\bar{w}}] \right|^2. 
   \label{scalar potential CP^2}
\end{eqnarray}
Recall that the background flux is given as 
\begin{equation}
\bar{F}_{mn}\ =\ f^a{}_{mn}T_a, 
\end{equation}
where $f^a{}_{mn}$ are the structure constants of $\mathfrak{g}=su(3)$, not of $\mathfrak{g}_{\rm YM}=su(4)$. 
In this case we find $\bar{F}_{zw}=0$ since the above choice of the generators for $su(3)$ 
in (\ref{definition coset CP^2}) and (\ref{definition coset CP^2-2})
gives $f^a{}_{zw}=0$. 

%\vspace{5mm}

Recall that the background gauge field $\bar{A}_\alpha$ is defined as 
\begin{equation}
\bar{A}_\alpha\ =\ e^a_\alpha T_a. 
\end{equation}
In the previous section, $e^a_\alpha$ gives a monopole configuration on $S^2$. 
Similarly, on $\mathbb{CP}^2$, $e^a_\alpha$ gives an instanton background. 
This can be deduced from the fact that $e^a_\alpha$ also gives the spin connection on $\mathbb{CP}^2$ as (\ref{spin connection}), and that the second Chern number of $\mathbb{CP}^2$ is non-zero \cite{Eguchi:1980jx}. 

In fact, this can be checked easily since the flux can be given expilcitly. 
We consider 
\begin{equation}
\bar{f}_{mn}\ :=\ f^a{}_{mn}t_a 
\end{equation}
as a flux of the ${\rm SU}(2)\times{\rm U}(1)$ gauge field on $\mathbb{CP}^2$. 
We notice that 
\begin{equation}
\bar{f}_{z\bar{z}}\ =\ -i\left(t_3+\frac32t_4\right), \hspace{5mm} \bar{f}_{w\bar{w}}\ =\ -
i\left(-t_3+\frac32t_4\right), \hspace{5mm} \bar{f}_{z\bar{w}}\ =\ -i\left(t_1+it_2\right) , \hspace{5mm} \bar{f}_{zw}\ =\ 0
\end{equation}
satisfy 
\begin{equation}
\bar{f}_{z\bar{z}}+\bar{f}_{w\bar{w}}\ =\ -3i\,t_4, \hspace{1cm} \bar{f}_{zw}\ =\ 0, \hspace{1cm} \bar{f}_{\bar{z}\bar{w}}\ =\ 0. 
\end{equation}
Note that $\bar{f}_{z\bar{z}}+\bar{f}_{w\bar{w}}$ is nonvanishing for the $u(1)$ part, and there is no 
$su(2)$ part. 
Since the instanton equation 
\begin{equation}
F_{mn}\ =\ -\frac12\epsilon_{mnkl}F_{kl}, 
\end{equation}
can be rewritten in terms of the complex coordinates as 
\begin{equation}
F_{z\bar{z}}+F_{w\bar{w}}\ =\ 0, \hspace{1cm} F_{zw}\ =\ 0, \hspace{1cm} F_{\bar{z}\bar{w}}\ =\ 0, 
\end{equation}
we find that the $su(2)$ part of the flux $\bar{f}_{mn}$ satisfies these instanton equation. 
Thus the $su(2)$ field strength is given by an $su(2)$ instanton configuration. 
But it does not always mean that the $\rm SU(4)$ configuration has a nonzero instanton number, as we will see later. 

%\vspace{5mm}

In the following, we consider two embeddings of ${\rm SU}(2)\times{\rm U}(1)$ into ${\rm SU}(4)$, and investigate the corresponding Higgs condensations. 
We will see that the topological nature of the background ${\rm SU}(4)$ flux plays an important role
in the gauge symmetry pattern when the symmetric Higgs acquires vev. 

%%%%%%%%%%%%%%%%%%%%%%%%%%%%%%%%%%%%%%%%
\vspace{5mm}

\subsection{Embedding of $H$$=$SU$(2)\times$U$(1)$ into $G_{\rm YM}$$=$SU$(4)$: Case 1}

\vspace{5mm}
Since $\rm SU(2)$ part of $H={\rm SU(2)\times U(1)}$ is embedded into $\rm SU(4)$ as (\ref{SU(2)embedding}), 
we choose an embedding of ${\rm U(1)} \subset H$ part. 
Our first choice for $T_4 \in u(1)$ is 
\begin{equation}
T_4\ =\ \frac13\left[
\begin{array}{cccc}
1 & 0 & 0 & 0 \\
0 & 1 & 0 & 0 \\
0 & 0 & -2 & 0 \\
0 & 0 & 0 & 0 
\end{array}
\right]. 
   \label{embedding CP^2 1}
\end{equation}
The background flux $\bar{F}_{mn}$ breaks the ${\rm SU}(4)$ gauge group to ${\rm U}(1)\times {\rm U}(1)$. 

Recall that the condition (\ref{condition symmetric Higgs}) for the symmetric Higgs fields is
\begin{equation}
[T_a,\phi_m]\ =\ if^n{}_{am}\phi_n
   \label{condition symmetric Higgs CP^2}
\end{equation}
where $T_a$ are Lie algebra generators of $H$.
In the previous section, this is a condition for a charge assigned to the components of $\phi_m$. 
For the $\mathbb{CP}^2$ case,
 $T_a$ form an $su(2)\times u(1)$ subalgebra of $su(4)$, 
 and the adjoint {representation ${\rm ad}_{\rm YM}$}, i.e., $\bf 15$ representation of $su(4)$,
 can be decomposed into irreducible representations of $su(2)\times u(1)$ as
\begin{equation}
\bf 15\ =\ 3_0\oplus2_1\oplus2_\frac13\oplus2_{-1}\oplus2_{-\frac13}\oplus1_\frac23\oplus1_{-\frac23}\oplus(1_0)^{\rm 2}.
\label{15representation}
\end{equation}
On the other hand, the same commutation relations are realized by the original $su(3)$ algebras,
\begin{equation}
[t_a,t_m]\ =\ if^n{}_{am}t_n ,
\end{equation}
and $t_m$ forms a set of irreducible representations of $su(2)\times u(1)$: 
\begin{equation}
R_t=\bf 2_1\oplus 2_{-1}.
\end{equation}
Then the symmetric Higgs fields $\phi_m$ satisfying (\ref{condition symmetric Higgs CP^2}) 
 can be obtained by those representations of $su(2)\times u(1)$ isomorphic to $R_t$
 in the irreducible decomposition of the adjoint representation $\bf 15$ of $su(4)$. 
From the definitions of $t_m$ in (\ref{definition coset CP^2-2}), we see that
 the symmetric Higgs fields are given by 
\begin{equation}
\phi_z\ =\ \left[
\begin{array}{cccc}
0 & 0 & \varphi & 0 \\
0 & 0 & 0 & 0 \\
0 & 0 & 0 & 0 \\
0 & 0 & 0 & 0 
\end{array}
\right], \hspace{1cm} 
\phi_w\ =\ \left[
\begin{array}{cccc}
0 & 0 & 0 & 0 \\
0 & 0 & \varphi & 0 \\
0 & 0 & 0 & 0 \\
0 & 0 & 0 & 0 
\end{array}
\right], \hspace{1cm} \phi_{\bar{s}}\ =\ (\phi_s)^\dag. 
\end{equation}
%\red{and its complex conjugate. }
Note that their non-zero components %of $\phi_z$ and $\phi_w$ 
must be the same in order to satisfy (\ref{condition symmetric Higgs CP^2}).

%The condition (\ref{condition symmetric Higgs CP^2}) can be understood in a group-theoretic manner. 
%We notice that $if^n{}_{am}$ give a representation $R_t$ of the $su(2)\times u(1)$ subalgebra realized on $t_m$. 
%Then, the condition (\ref{condition symmetric Higgs CP^2}) means that the symmetric Higgs fields correspond to those representations of $su(2)\times u(1)$ isomorphic to $R_t$ which appears in the irreducible decomposition of the adjoint representation $\bf 15$ of $su(4)$. 
%In fact, we have 
%\begin{equation}
%\bf 15\ =\ 3_0\oplus2_1\oplus2_\frac13\oplus2_{-1}\oplus2_{-\frac13}\oplus1_\frac23\oplus1_{-\frac23}\oplus(1_0)^{\rm 2}
%\end{equation}
%and $R_t=\bf 2_1\oplus 2_{-1}$. 

The scalar potential $V(\phi)$ is then given by
\begin{equation}
V(\phi)\ =\ \frac34\left( 1-|\varphi|^2 \right)^2. 
\end{equation}
Therefore, the symmetric Higgs field $\varphi$ acquires vev at $\varphi=1$ up to a gauge transformation. 
This breaks the residual ${\rm U}(1)\times {\rm U}(1)$ gauge symmetry to ${\rm U}(1)$, and within the
analysis of the  low lying modes in the coset space compactification,
the symmetry breaking pattern would be 
 \begin{equation}
  G_{\rm YM}={\rm SU}(4) \xrightarrow{{\rm background}\ {\rm flux}} {\rm U}(1) \times {\rm U}(1)
  \xrightarrow{{\rm Higgs} \ {\rm vev}} {\rm U}(1) \  ?
\end{equation}

%\vspace{5mm}

We have found that the symmetric Higgs vacuum attains the global minimum of the scalar potential. 
This implies the stability of the vacuum, and the restoration of gauge symmetry when the Higgs acquires vev. 
In fact, this turns out to happen in more general situations \cite{Harnad:1980fz,Chapline:1980mr}. 
This will become apparent when we reconsider the above calculations as follows
to elaborate the reason why $|\varphi|=1$ attains the global minimum of the potential in the present setup.
We have considered the embedding (\ref{SU(2)embedding}) and (\ref{embedding CP^2 1}), 
which can be generalized to the other generators as 
\begin{equation}
T_m\ :=\ \left[
\begin{array}{cccc}
 & & & 0 \\
 & t_m & & 0 \\
 & & & 0 \\
 0 & 0 & 0 & 0
\end{array}
\right].
\label{remaining-su3-generators}
\end{equation}
Then, $T_a$ and $T_m$ form an $su(3)$ subalgebra embedded into the $3\times3$ upper-left block of $su(4)$. 
%This implies that they satisfy the same commutation relations as those for $t_a$ and $t_m$. 
The condition for the symmetric Higgs fields is given by (\ref{condition symmetric Higgs CP^2}),
%\begin{equation}
%[T_a,\phi_m]\ =\ if^n{}_{am}\phi_n, 
%\end{equation}
and the condition for the vanishing scalar potential in (\ref{scalar potential algebraic}) is written as
\begin{equation}
[\phi_m,\phi_n]\ =\ if^a{}_{mn}T_a. 
\end{equation}
Comparing these two conditions, 
we find that $\phi_m=T_m$ is a soution for both conditions since they become nothing but a part of the commutation relations of $su(3)$. 
This is the reason why the symmetric Higgs vev $\varphi=1$ attains the global minimum of the scalar potential. 
Now, it is clear that this phenomenon always happens for a general coset space $G/H$, if we choose an embedding of $H$ into $G_{\rm YM}$ which is induced by an embedding of $G$ into $G_{\rm YM}$. 
In fact, we have already observed this phenomenon in section \ref{S^2 1} for  $G/H=S^2$. 

%\vspace{5mm}

At this point, one might be puzzled by the fact that the symmetric Higgs vacuum attains the global minimum of the scalar potential, especially when one remembers that the background gauge filed consists of an instanton configuration. 
On the one hand, the Higgs condensation is nothing but a continuous deformation of the gauge field configuration on $\mathbb{CP}^2$. 
On the other hand, the vanishing potential implies that the gauge field configuration is just a pure gauge. 
This looks contradicting to the topologically non-trivial nature of the instanton configuration. 
The resolution of this puzzle comes from the fact that the $su(2)$ instanton is embedded into $su(4)$ {\it with} a $u(1)$ flux,
and the instanton number is cancelled between the $su(2)$ and $u(1)$ parts. 
Indeed, we can calculate the instanton number of the background gauge field $\bar{A}_\alpha$ for the $su(4)$ gauge field explicitly,
and find 
\begin{eqnarray}
\frac12\epsilon_{mnkl}{\rm Tr}\,F_{mn}F_{kl} 
&=& {\rm Tr}\left( -F_{z\bar{z}}F_{w\bar{w}}+F_{zw}F_{\bar{z}\bar{w}}-F_{z\bar{w}}F_{\bar{z}w} \right) \nonumber \\
&=& {\rm Tr}\left( -T_3T_3+\frac94T_4T_4-T_1T_1-T_2T_2 \right) \nonumber \\
&=& 0. 
\end{eqnarray}
The $(T_4)^2$ part is a contribution from the $u(1)$,    and 
we conclude that the gauge field configuration before the Higgs condensation has zero instanton number. 
This is compatible with the fact that the gauge field configuration at the symmetric Higgs vacuum is trivial. 

%\vspace{5mm}

Finally, let us check whether there are 15 massless vector fields at the symmetric Higgs vacuum. 
The massless conditions in this case are 
\begin{equation}
\bar{D}_za_\mu+i[T_z,a_\mu]\ =\ 0, \hspace{1cm} \bar{D}_wa_\mu+i[T_w,a_\mu]\ =\ 0,
   \label{condition massless vector field CP^2}
\end{equation}
 where $T_z$ and $T_w$ are defined in (\ref{remaining-su3-generators})
 with $m=z$ and $w$, respectively.
Since $T_a$ and $T_m$ form an $su(3)$ subalgebra of $su(4)$, as mentioned above, it is convenient to decompose these massless conditions according to the irreducible decomposition of $\bf 15$ of $su(4)$ to $\bf 8\oplus3\oplus\bar{3}\oplus1$ of the $su(3)$ subalgebra. 
By rearranging the components $a_{\mu,ij}$ into the corresponding vectors $a^{(R)}_\mu$ with $R={\bf 8,3,\bar{3},1}$, we obtain 
\begin{equation}
\bar{D}_za^{(R)}_\mu\ =\ -iT^{(R)}_z\cdot a^{(R)}_\mu, \hspace{1cm} \bar{D}_wa^{(R)}_\mu\ =\ -iT^{(R)}_w\cdot a^{(R)}_\mu, 
   \label{decomposition massless condition}
\end{equation}
where $T^{(R)}_m$ are generators in the representation $R$. 

As reviewed in Appendix \ref{mode functions appendix}, 
each component of the vectors $a^{(R)}_\mu$ on $\mathbb{CP}^2$ can be expanded 
by the complete set of functions $f^R_{IJ}(y)$ where $R$ runs over all representations of ${\rm SU}(3)$ and $I,J$ are the indices for the representation $R$, that is, they run from 1 to $\dim R$. 
Recall that the representation $R$ and one of the indices $I$ are constrained by a condition of what kind of 
representation of $H$ we are investigating on the coset space $G/H$. 
In the previous section, we used one index $m$ of the mode functions $f^j_{mm'}(y)$ to indicate its $T$-charges $q$,
and $j$ is constrained so that the representation contains the desired value of $m=q.$ 
Similarly, in this case, we use one index $I$ of $f^R_{IJ}(y)$ to indicate its $su(2)\times u(1)$ representation. 
Namely, if the irreducible decomposition of $R$ has a representation $r$ of $su(2)\times u(1)$, then $f^R_{iJ}(y)$ contributes to the expansion of a field in the representation $r$ of $su(2)\times u(1)$, where $i$ runs from 1 to $\dim r$. 
Therefore, the expansion of a field $\chi_i(y)$ in the representation $r$ is given as 
\begin{equation}
\chi_i(y)\ =\ \sum_{r\subset R}\sum_{J=1}^{\dim R}c^R_{J}f^R_{iJ}(y), 
\end{equation}
where the first summation is over the representations $R$ of $su(3)$ whose irreducible decomposition with respect to $su(2)\times u(1)$ has $r$. 
If the decomposition of $R$ contains several irreducibe representations each of which is isomorphic to $r$, then the multiplicity is also taken into account in the sum. 

The $su(2)\times u(1)$ representations for $a^{(R)}_\mu$ can be found by further decomposition of $R$ with respect to $su(2)\times u(1)$ subalgebra of $su(3)$. 
Explicitly, 
\begin{eqnarray}
\bf 8 &=& \bf 3_0\oplus2_1\oplus2_{-1}\oplus 1_0, \\
\bf 3 &=& \bf 2_{\frac13}\oplus1_{-\frac23}, \\
\bf \bar{3} &=& \bf 2_{-\frac13}\oplus1_{\frac23}, \\
\bf 1 &=& \bf 1_0. 
\end{eqnarray}

The action of the covariant derivatives $\bar{D}_z, \bar{D}_w$ on the mode functions $f^R_{iJ}(y)$ is again given by the multipication of $T^{(R)}_z, T^{(R)}_w$ from the left. 
Therefore, the massless condition is again reduced to the requirement that the adjoint action of $T_z, T_w$ due to the symmetric Higgs vev has the same effect on $a_\mu$ as the action of $T^{(R)}_z,T^{(R)}_w$ on the mode functions. 

In the present case, the solution to the massless conditions is almost obvious. 
For example, let $i_1,i_2$ be indices for the representations $\bf 2_{\frac13},1_{-\frac23}$, respectively. 
Then 
\begin{equation}
a^{({\bf 3})}_\mu (x,y)\ =\ \left[
\begin{array}{c}
\sum_{J=1}^3 c^{\bf 3}_{J} (x) f^{\bf 3}_{i_1J}(y) \\ [2mm] 
\sum_{J=1}^3c^{\bf 3}_{J} (x)   f^{\bf 3}_{i_2J}(y)
\end{array}
\right]
\end{equation}
is the solution for $R={\bf 3}$. 
Note that the expansion coefficients in the first and the second rows are the same. 
The solutions for the other $R$ can be obtained similarly. 
They give us {\bf 15 massless vector fields}, as expected. 
Thus, the gauge symmetry is restored and we have the symmetry breaking-restoration pattern
\begin{equation}
  G_{\rm YM}={\rm SU}(4) \xrightarrow{{\rm background}\ {\rm flux}} {\rm U}(1) \times {\rm U}(1)
  \xrightarrow{{\rm Higgs} \ {\rm vev}} {\rm SU}(4). \  
\end{equation}

 %%%%%%%%%%%%%%%%%%%%%%%%%%%%%%%%%%%%%%%
 %%%%%%%%%%%%%%%%%%%%%%%%%%%%%%%%%%%%%%%
 %%%%%%%%%%%%%%%%%%%%%%%%%%%%%%%%%%%%%%%
 %%%%%%%%%%%%%%%%%%%%%%%%%%%%%%%%%%%%%%%

\vspace{5mm}

\subsection{Embedding of $H$$=$SU$(2)\times$U$(1)$ into $G_{\rm YM}$$=$SU$(4)$: Case 2}
\label{CP^2 with instanton}

\vspace{5mm}
Let us consider a different embedding of $H$ into $G_{\rm YM}$. 
The SU(2) part of $H={\rm SU(2)\times U(1)}$
 is embedded into $G_{\rm YM}={\rm SU(4)}$ as (\ref{SU(2)embedding}). 
Our second choice of the U(1) part, $T_4$, into $\rm SU(4)$ is 
\begin{equation}
T_4\ =\ \frac12\left[
\begin{array}{cccc}
1 & 0 & 0 & 0 \\
0 & 1 & 0 & 0 \\
0 & 0 & -1 & 0 \\
0 & 0 & 0 & -1 
\end{array}
\right]. 
\end{equation}
The corresponding background flux breaks the ${\rm SU}(4)$ gauge group to ${\rm SU}(2)\times {\rm U}(1)$. 
The symmetric Higgs fields are then given by
\begin{equation}
\phi_z\ =\ \left[
\begin{array}{cccc}
0 & 0 & \varphi_1 & \varphi_2 \\
0 & 0 & 0 & 0 \\
0 & 0 & 0 & 0 \\
0 & 0 & 0 & 0 
\end{array}
\right], \hspace{1cm} 
\phi_w\ =\ \left[
\begin{array}{cccc}
0 & 0 & 0 & 0 \\
0 & 0 & \varphi_1 & \varphi_2 \\
0 & 0 & 0 & 0 \\
0 & 0 & 0 & 0 
\end{array}
\right]. 
\label{symmetric Higgs cp2-2}
\end{equation}
Note that the irreducible decomposition of $\bf 15$ is now 
\begin{equation}
\bf 15\ =\ 3_0\oplus (2_1)^{\rm 2}\oplus (2_{-1})^{\rm 2}\oplus (1_0)^{\rm 4}
\end{equation}
due to the different choice of $T_4$. 
Thus there are two $R_t =\bf 2_1\oplus2_{-1}$ representations in $\bf 15$, and 
we have two symmetric Higgs. 
The components $\varphi_s$ $(s=1,2)$ form a doublet of the ${\rm SU}(2)$ gauge group. 

%\vspace{5mm}

The calculation of the scalar potential for $\varphi_s$ is rather complicated if we just insert the above expressions into (\ref{scalar potential CP^2}). 
It is better to keep track of the residual gauge invariance. 
For this purpose, we rewrite a part of the 
commutation relations of $su(4)$ relevant for calculating $V(\phi)$ so that the residual ${\rm SU}(2)$ gauge symmetry becomes manifest. 

Let us explicitly write some generators of $su(4)$ other than $T^a$. 
First, we define
\begin{equation}
\tilde{T}_i\ :=\ \left[
\begin{array}{cc}
\begin{array}{cc}
0 & 0 \\
0 & 0
\end{array}
& \hspace{-3mm}
\begin{array}{cc}
0 & 0 \\
0 & 0
\end{array}
\\
\begin{array}{cc}
0 & 0 \\
0 & 0 
\end{array}
 & \hspace{-2mm} \frac12\sigma_i
\end{array}
\right]. 
\end{equation}
These three $\tilde{T}_i$ and $T_4$ correspond to the generators of the residual gauge symmetry SU(2)$\times$U(1). 
The generators in the off-diagonal components are relabeled as 
\begin{equation}
T_z^{1}\ :=\ T_z, \hspace{5mm} T_w^{1}\ :=\ T_w, \hspace{5mm} T_z^{2}\ :=\ \left[
\begin{array}{cccc}
0 & 0 & 0 & 1 \\
0 & 0 & 0 & 0 \\
0 & 0 & 0 & 0 \\
0 & 0 & 0 & 0 
\end{array}
\right], \hspace{5mm} T_w^{2}\ :=\ \left[
\begin{array}{cccc}
0 & 0 & 0 & 0 \\
0 & 0 & 0 & 1 \\
0 & 0 & 0 & 0 \\
0 & 0 & 0 & 0 
\end{array}
\right], 
\end{equation}
 where $T_z$ and $T_w$ are defined in (\ref{remaining-su3-generators}) with $m=z$ and $w$, respectively,
 and $T_{\bar{s}}^{\alpha}:=(T_s^{\alpha})^\dag$ with $s=z,w$. 
Then, the symmetric Higgs field $\phi_s$ in (\ref{symmetric Higgs cp2-2})
can be written in terms of these generators as 
\begin{equation}
\phi_s\ = \sum_{\alpha=1,2} \varphi_{\alpha}T_s^{\alpha}, \hspace{1cm} 
\phi_{\bar{s}}\ =\sum_{\alpha=1,2} \varphi_\alpha^\dag T_{\bar{s}}^\alpha. 
\end{equation}
The relevant commutation relations for calculating $V(\phi)$ are 
\begin{equation}
[T_s^\alpha,T_{\bar{t}}^\beta]\ =\ (\sigma_i)_{\bar{t}s}\delta^{\alpha\beta}T_i+\delta_{\bar{t}s}\delta^{\alpha\beta}T_4-\delta_{\bar{t}s}(\sigma_i)^{\alpha\beta}\tilde{T}_i, 
\end{equation}
where $(\sigma_i)_{\bar{z}z}:=(\sigma_i)_{11}$ etc. 
Then we obtain 
\begin{equation}
[\phi_s,\phi_{\bar{t}}]\ =\ |\varphi|^2(\sigma_i)_{\bar{t}s}T_i+|\varphi|^2\delta_{\bar{t}s}T_4-\delta_{\bar{t}s}(\varphi\sigma_i\varphi^\dag)\tilde{T}_i, 
\end{equation}
where $|\varphi|^2:=|\varphi_1|^2+|\varphi_2|^2$. 

We also need to rewrite the commutation relation for $t_a$ in $su(3)$ as 
\begin{equation}
[t_s,t_{\bar{t}}]\ =\ (\sigma_i)_{\bar{t}s}t_i+\frac32\delta_{\bar{t}s}t_4. 
\end{equation}
Then the background flux can be written as 
\begin{equation}
\bar{F}_{s\bar{t}}\ =\ -i\left[ (\sigma_i)_{\bar{t}s}T_i+\frac32\delta_{\bar{t}s}T_4 \right]. 
\end{equation}
Note that $\bar{F}_{st}$ and $\bar{F}_{\bar{s}\bar{t}}$ vanish. 

By using the above expressions, we find 
\begin{eqnarray}
\bar{F}_{z\bar{z}}+i[\phi_z,\phi_{\bar{z}}] 
&=& i(|\varphi|^2-1)T_3+i\left( |\varphi|^2-\frac32 \right)T_4-i(\varphi\sigma_i\varphi^\dag)\tilde{T}_i, 
   \label{flux including Higgs vev 1} \\
\bar{F}_{w\bar{w}}+i[\phi_w,\phi_{\bar{w}}]  
&=& -i(|\varphi|^2-1)T_3+i\left( |\varphi|^2-\frac32 \right)T_4-i(\varphi\sigma_i\varphi^\dag)\tilde{T}_i, 
   \label{flux including Higgs vev 2} \\
\bar{F}_{z\bar{w}}+i[\phi_z,\phi_{\bar{w}}] 
&=& i(|\varphi|^2-1)(T_1-iT_2). 
   \label{flux including Higgs vev 3}
\end{eqnarray}
Finally, the potential $V(\phi)$ turns out to be 
\begin{eqnarray}
V(\phi) 
&=& \frac14\left[ \frac12\left( 1-|\varphi|^2 \right)^2+\left( \frac32-|\varphi|^2 \right)^2+\frac12(\varphi\sigma_i\varphi^\dag)^2 \right]+\frac14\left( 1-|\varphi|^2 \right)^2 \nonumber \\
&=& \frac34\left( 1-|\varphi|^2 \right)^2+\frac3{16}. 
\end{eqnarray}
The symmetric Higgs doublet therefore acquires the vev 
\begin{equation}
\varphi_1\ =\ 1, \hspace{1cm} \varphi_2\ =\ 0, 
\end{equation}
up to a global gauge transformation. 
The non-zero value of the scalar potential at the symmetric Higgs vacuum implies that there remains a flux at this vacuum, which suggests that the gauge symmetry at the symmetric Higgs vacuum must be smaller than ${\rm SU}(4)$. 

%\vspace{5mm}

Now, we count the number of massless vector fields at the symmetric Higgs vacuum. 
The massless conditions are given as (\ref{condition massless vector field CP^2}), exactly the same condition we discussed in the previous section. 
Therefore, we can employ the decomposition (\ref{decomposition massless condition}) again. 

Let us consider 
\begin{equation}
\bar{D}_za^{(\bf 3)}_\mu\ =\ -iT^{(\bf 3)}_z\cdot a^{(\bf 3)}_\mu, \hspace{1cm} \bar{D}_wa^{(\bf 3)}_\mu\ =\ -iT^{(\bf 3)}_w\cdot a^{(\bf 3)}_\mu. 
   \label{massless condition CP^2-2}
\end{equation}
Recall that $a^{(\bf 3)}_\mu$ is formed from the components $a_{\mu,14}$, $a_{\mu,24}$ and $a_{\mu,34}$. 
In terms of the $su(2)\times u(1)$ subgroup generated by $T_a$, this consists of the representations $\bf 2_1\oplus 1_0$. 
As explained in Appendix \ref{mode functions appendix}, 
they are expanded by the mode functions $f^R_{IJ}(y)$, where $R$ is a representation of $su(3)$ whose irreducible decomposition contains $\bf 2_1$ or $\bf 1_0$. 
An important point is that this irreducible decomposition must be considered with respect to the $su(2)\times u(1)$ subgroup (\ref{definition coset CP^2}), not to any $su(4)$ embeddings. 
According to this, the representation $\bf 3$ of the $su(3)$ is decomposed as $\bf 2_{\frac13}\oplus 1_{-\frac23}$. 
This implies that the mode functions $f^{\bf 3}_{IJ}(y)$ do not contribute to the expansion of $a^{(\bf 3)}_\mu$. 
Instead, other mode functions, for example, $f^{\bf 8}_{IJ}(y)$ contribute to the expansion since $\bf 8$ is decomposed as $\bf 3_0\oplus 2_1\oplus 2_{-1}\oplus 1_0$. 
Then, the covariant derivatives $\bar{D}_z, \bar{D}_w$ can be converted to $T^{(\bf 8)}_z, T^{(\bf 8)}_w $ but never to $T^{(\bf 3)}_z, T^{(\bf 3)}_w$. 
By this reason, we conclude that the massless conditions (\ref{massless condition CP^2-2}) do not have solutions. 
Similarly, the conjugate components $a^{(\bf \bar{3})}_\mu$ give us only massive vector fields. 

The conditions for the other two representations $\bf 8$ and $\bf 1$
 turn out to give us $8+1$ massless vector fields. 
A natural guess is that the emergent gauge symmetry at the symmetric Higgs vacuum would be $\rm SU(3)\times U(1)$. 
\begin{equation}
  G_{\rm YM}={\rm SU}(4) \xrightarrow{{\rm background}\ {\rm flux}} {\rm SU}(2) \times {\rm U}(1)
%  {\color{blue} \ ({\rm SU}(2) \times {\rm U}(1) ?)}
  \xrightarrow{{\rm Higgs} \ {\rm vev}} {\rm SU}(3)\times{\rm U}(1). \  
\end{equation}
This should be confirmed by a further analysis. 

%\vspace{5mm}

At the symmetric Higgs vacuum obtained above, the scalar potential is non-vanishing. 
This is a similar situation as the one  discussed in section \ref{S^2 2}. 
Interestingly, we can show that the symmetric Higgs vacuum in this section is stable,
and the topological nature of the background gauge field plays an important role. 
In the following, we see 
that the background gauge field $\bar{A}_\alpha$ before the Higgs condensation has a non-zero instanton number,
which is unchanged by any continuous deformation of the gauge field.
The non-zero instanton number is obtained in this setup since
the embedding $T_4$ of $t_4$ is different from the one we chose in the previous section. 
Let us
calculate the instanton number  for the gauge field configuration, including the symmetric Higgs vev,  given by
\begin{equation}
F_{mn}\ =\ \bar{F}_{mn}+i[\phi_m,\phi_n], \hspace{1cm} \phi_m\ =\ \varphi_\alpha T^\alpha_m. 
\end{equation}
Their explicit forms are given in (\ref{flux including Higgs vev 1})(\ref{flux including Higgs vev 2})(\ref{flux including Higgs vev 3}). 
We find 
\begin{eqnarray}
\frac12\epsilon_{mnkl}{\rm Tr}\,F_{mn}F_{kl} 
&=& \frac12\left( |\varphi|^2-1 \right)^2-\left( |\varphi|^2-\frac32 \right)^2-\frac12\left( \varphi\sigma_i\varphi^\dag \right)^2+\left( |\varphi|^2-1 \right)^2 \nonumber \\
&=& -\frac34. 
\end{eqnarray}
The scalar potential is bounded by this instanton number density as 
\begin{eqnarray}
V(\phi) 
&=& \frac18{\rm Tr}\,\left( F_{mn}+\frac12\epsilon_{mnkl}F_{kl} \right)^2-\frac18\epsilon_{mnkl}{\rm Tr}\,F_{mn}F_{kl} \nonumber \\
&\ge& \frac3{16}. 
\end{eqnarray}
This shows that the symmetric Higgs vacuum attains the global minimum of the scalar potential in a given topological sector of the gauge field. 
In other words, we can say that the non-trivial topological nature of the original background gauge field $\bar{A}_\alpha$ stabilizes the non-trivial symmetric Higgs vacuum. 

%\vspace{5mm}

This example tells us the geometric picture of the Higgs condensation realized in the Kaluza-Klein reduction of Yang-Mills theory. 
Since the Higgs fields come from some components of the gauge field, the Higgs condensation is nothing but a particular continuous deformation of the gauge field. 
The condensation occurs in order to minimize the ``Euclidean action'' 
\begin{equation}
S_E\ :=\ \frac14\int dv\, {\rm Tr}\,F_{mn}F_{mn}. 
\end{equation}
The configuration space of the gauge field is divided into various components according to certain topological invariants. 
The vacuum corresponds to the global minimum of $S_E$ in a given component. 
The global minimum may or may not be attained by the condensation of symmetric Higgs fields, according to the situation. 

It is interesting to notice that the gauge field configuration $F_{mn}$ including the symmetric Higgs vev becomes exactly an instanton configuration at the symmetric Higgs vacuum. 
This means that we find an explicit construction of the instanton solution on $\mathbb{CP}^2$. 
It is interesting to clarify how general this construction is.

%%%%%%%%%%%%%%%%%%%%%%%%%%%%%%%%%%%%%%%
 %%%%%%%%%%%%%%%%%%%%%%%%%%%%%%%%%%%%%%%

\vspace{1cm}

\section{Conclusions}

\vspace{5mm}

Higher-dimensional gauge theories with non-trivial fluxes of 
the background gauge fields in the compact space have been investigated
as phenomenological models of gauge symmetry breaking. 
The background gauge fluxes  explicitly break some of the original gauge symmetries, 
and often provide  tachyonic scalar fields whose vacuum expectation value further
breaks the remaining gauge symmetries  in the four dimensional effective field theory. 
Thus,  this kind of models are often utilized for dynamical generation of Higgs potential. 
In their investigations, 
 we usually study four-dimensional effective field theories 
 by keeping only the light fields and  neglecting other massive modes in the Kaluza-Klein tower, 
 since we are interested in physics in the lower energy scale than that of the compact space.  

In this paper, we revisit  such higher dimensional models
including all the massive Kaluza-Klein modes to investigate their roles in the gauge symmetry breaking.
The inclusion of higher modes will be important  since 
the scale  of the vacuum expectation value of the Higgs field is usually given by the same scale 
as the mass of the Kaluza-Klein modes. 
Indeed both of them are given by the scale of the compact space. 
What we have shown in the present paper is 
 that, when the Higgs acquires vacuum expectation value at the minimum of the potential, 
some of the originally massive vector fields in the Kaluza-Klein tower become massless
and the corresponding gauge symmetries are restored. 
If we restricted ourselves to consider only the light modes, 
we would conclude that the gauge symmetry, which remains to be unbroken by the gauge flux, 
 would be further broken by the Higgs field. But, if we consider all the Kaluza-Klein modes, 
 the symmetries, on the contrary, are restored even to the full set of the original gauge symmetries. 
 When the massive gauge field becomes massless, it will provide an additional massless scalar field to 
 the massless gauge field. 
Possible candidates for such massless scalar fields are, for example,
 instanton moduli in the model discussed in section \ref{CP^2 with instanton}.
It is interesting to see whether the scalar field acquires mass
 due to the radiative corrections or remains massless against perturbations. 

We have studied two classes of the compact space, $S^2$ and $\mathbb{CP}^2$.
Both of them are coset spaces $G/H$, and  due to the beautiful group theoretical structures of the coset space, 
we have succeeded to analyze the mass spectrum even when the Higgs acquires vacuum expectation value. 
In some cases, all the gauge symmetries are recovered, and in other cases, 
only a part of them are recovered. 
The vacuum is shown to be stable even in the latter cases.
The stability is related to the topological structures of the gauge field configurations. 
In cases when  the background gauge field configurations are topologically trivial,
the original gauge symmetries are completely restored at the global minimum of the potential. 
On the other hand, 
 in cases when the background gauge configurations are topologically non-trivial, 
 or when they have some conserved topological numbers,
 the original gauge symmetries can be only partially restored. 
The background gauge field configurations including the Higgs fields in the true vacua
are described by new topologically non-trivial configurations with the same topological numbers.

We have developed a group-theoretic technique for analyzing the number of massless vector fields at the symmetric Higgs vacuum. 
It is reasonalble to expect that the technique can be extended to obtain the mass formulas applicable also to massive vector fields and scalar fields. 
A key result which enables the technique to work is that the symmetric Higgs vacuum expectation values coincide with some generators of the gauge group $G_{\rm YM}$. 
It is interesting to clarify whether this happens in more general models. 
A detailed understanding of the structure of the scalar potential will help us to gain insights on this issue. 

An interesting observation we made is that the Higgs vacuum in a model invetigated in section \ref{CP^2 with instanton} corresponds to an instanton solution on $\mathbb{CP}^2$. 
The solution can be given quite explicitly from the Maurer-Cartan 1-form on $G$. 
It is curious to see whether this can be a general method of constructing instanton solutions on coset spaces. 
Our calculation suggests that the different embeddings of $H$ into $G_{\rm YM}$ would give us instanton solutions with different instanton numbers. 

The effects of gravity are neglected in the investigations of the present paper. 
One of the original physics target of the coset compactifications
 is the stabilization of the compact space with gravity,
called, ``spontaneous compactification'',
in which the background flux in the coset space is 
a classical solution of the gauge field equations as well as the Einstein equation. 
It will be interesting to investigate the stability and the pattern of gauge symmetry breaking including the fluctuations of gravity. 

The Einstein-Yang-Mills theory naturally appears as the bosonic part of the low energy effective theory of the heterotic string theory. 
A Higgs field which appears in this context corresponds to a closed string tachyon, whose condensation is an interesting research subject in theoretically as well as phenomenologically. 
The condensation of closed string tachyon was discussed,
 for example, in \cite{Adams:2001sv,Suyama:2001bn}.
Since the Einstein-Yang-Mills theory can be regarded as a truncation of string field theory, it is a natural arena for discussing the Higgs (closed string tachyon) condensation. 
It is interesting if there would exist an endpoint of the condensation which is stabilized due to a topological reason.

\vspace{2cm}

%{\bf \LARGE Acknowledgements}
\section*{Acknowledgements} 
The work is supported in part by Grants-in-Aid for Scientific Research
 No.16K05329, No.18H03708 and No.19K03851
 from the Japan Society for the Promotion of Science. 
%

%%%%%%%%%%%%%%%%%%%%%%%%%%%%%%%%%%%%%%%
 %%%%%%%%%%%%%%%%%%%%%%%%%%%%%%%%%%%%%%%
%%%%%%%%%%%%%%%%%%%%%%%%%%%%%%%%%%%%%%%
 %%%%%%%%%%%%%%%%%%%%%%%%%%%%%%%%%%%%%%%

\appendix

\vspace{2cm}

\section{Coset spaces}   \label{coset space appendix}

\vspace{5mm}

In this appendix, we summarize some mathematical basics of coset spaces. 
For further details, see, e.g., \cite{Kapetanakis:1992hf} or \cite{Kobayashi-Nomizu}. 

%\vspace{5mm}

Let $G$ be a group, and  $H$ be a subgroup of $G$. 
For an element $g$ of $G$, we define a subset $gH$ of $G$ as 
\begin{equation}
gH\ :=\ \{\,gh\,|\,h\in H\,\}.
\end{equation}
The set of such subsets is denoted by $G/H$. 

Suppose that $g$ belongs to both $g_1H$ and $g_2H$ in $G/H$. 
Then, $g$ can be written as $g=g_1h_1$ and $g=g_2h_2$, which imply $g_1=g_2h_2h_1^{-1}$. 
This then implies $g_1H=g_2H$. 
In other words, $g_1H$ and $g_2H$ in $G/H$ are either equal or disjoint as subsets of $G$. 

We choose one representation $g_iH$ for each element of $G/H$. 
Then, the set $\{g_i\}$ of elements of $G$ is in one-to-one correspondence to $G/H$. 
The elements $g_i$ are called a set of representatives for $G/H$. 
In terms of the representatives, $G$ can be written as 
\begin{equation}
G\ =\ \bigcup_i g_iH, 
   \label{coset decomposition}
\end{equation}
where $g_iH$ and $g_jH$ are disjoint to each other if $g_i\ne g_j$. 

%\vspace{5mm}

We can perform the same construction when $G$ is a Lie group and $H$ is its closed subgroup. 
Note that $H$ is always a Lie subgroup of $G$. 
In this case, $G/H$ is known to be a smooth manifold. 
This is called a coset space. 
In mathematics literature, this is also called a homogeneous space. 
If the dimensions of $G$ and $H$ are $d_G$ and $d_H$, respectively, then the dimension of $G/H$ is $d:=d_G-d_H$. 
A set of representatives for $G/H$ is regarded as a $G$-valued function on $G/H$. 
For a given local coordinate patch $U$ of $G/H$, the representatives corresponding to points in $U$ can be chosen such that they are given by a smooth function $g(y)$ on $U$ where $y^\alpha$ are coordinates on $U$. 

In this paper, we focus our attention on a particular class of coset spaces, known as symmetric coset spaces. 
They are characterized by the structures of the Lie algebras $\mathfrak{g}$ and $\mathfrak{h}$ of $G$ and $H$, respectively. 
Let $t_a$ be a set of generators of $\mathfrak{h}$ where $a$ runs from 1 to $d_H$. 
Their commutation relations are 
\begin{equation}
[t_a,t_b]\ =\ if^{c}{}_{ab}t_c, 
\end{equation}
since $\mathfrak{h}$ is a subalgebra of $\mathfrak{g}$. 
Let $t_m$ be a set of generators of $\mathfrak{g}$ other than $t_a$ where $m$ runs from 1 to $d$. 
We require that the other commutation relations of $\mathfrak{g}$ are of the form 
\begin{equation}
[t_a,t_m]\ =\ if^n{}_{am}t_n, \hspace{1cm} [t_m,t_n]\ =\ if^a{}_{mn}t_a. 
\end{equation}
A coset space $G/H$ in which the Lie algebras $\mathfrak{g,h}$ satisfy this condition is said to be symmetric. 
This condition can be understood as follows. 
For a symmetric coset space, we can assign a ``parity'' for generators such that $t_a$ has $+1$ and $t_m$ has $-1$. 
In sections \ref{condensation  S^2} and \ref{condensation CP^2}, we consider particular symmetric coset spaces, namely $S^2$ and $\mathbb{CP}^2$. 

%\vspace{5mm}

First, let us describe $S^2$ as a symmetric coset space. 
For this purpose, we regard $S^2$ as $\mathbb{CP}^1$. 
A point in $\mathbb{CP}^1$ is represented by a pair of complex numbers $(c_1,c_2)\in\mathbb{C}^2$ with $(c_1,c_2)\ne(0,0)$.
Two such pairs 
$(c_1,c_2)$ and $(c_1',c_2')$ correspond to the same point of $\mathbb{CP}^1$ if and only if there exists a non-zero $\lambda\in\mathbb{C}$ such that 
\begin{equation}
(c_1',c_2')\ =\ (\lambda c_1,\lambda c_2)
\end{equation}
is satisfied. 
Using this ambiguity, we can always choose the pair $(c_1,c_2)$ such that they satisfy $|c_1|^2+|c_2|^2=1$. 
Any pair $(c_1,c_2)$ satisfying $|c_1|^2+|c_2|^2=1$ can be written as 
\begin{equation}
\left[
\begin{array}{c}
c_1 \\
c_2
\end{array}
\right]\ =\ U\left[
\begin{array}{c}
1 \\
0
\end{array}
\right], \hspace{1cm} U\in {\rm SU}(2). 
\end{equation}
A pair $(c_1,c_2)$ represents the same point in $\mathbb{CP}^1$ as $(1,0)$ if and only if $U$ is of the form 
\begin{equation}
U\ =\ \left[
\begin{array}{cc}
e^{i\varphi} & 0 \\
0 & e^{-i\varphi}
\end{array}
\right]. 
\end{equation}
Such elements form a ${\rm U}(1)$ subgroup of ${\rm SU}(2)$. 
We have found that $S^2=\mathbb{CP}^1$ can be written as the coset space ${\rm SU}(2)/{\rm U}(1)$. 
In this case, $t_a$ corresponds to $\frac12\sigma_3$ and $t_m$ correspond to $\sigma_\pm$. 
Their commutation relations show that ${\rm SU}(2)/{\rm U}(1)$ is symmetric. 

Next, consider $\mathbb{CP}^2$. 
This is a straightforward generalization of the $\mathbb{CP}^1$ case. 
Any point of $\mathbb{CP}^2$ corresponds to a triple $(c_1,c_2,c_3)\in\mathbb{C}^3$ satisfying $|c_1|^2+|c_2|^2+|c_3|^2=1$, which can be written as 
\begin{equation}
\left[
\begin{array}{c}
c_1 \\
c_2 \\
c_3
\end{array}
\right]\ =\ U\left[
\begin{array}{c}
1 \\
0 \\
0
\end{array}
\right], \hspace{1cm} U\in {\rm SU}(3). 
\end{equation}
The multiplication by $U$ does not change the point in $\mathbb{CP}^2$ if and only if  $U$ is of the form 
\begin{equation}
\left[
\begin{array}{cc}
e^{2i\varphi} & 0\ \ 0  \\
\begin{array}{c}
0 \\
0
\end{array} & e^{-i\varphi}U \\
\end{array}
\right], \hspace{1cm} U\in \rm SU(2). 
\end{equation}
The matrices of the form consist of an $\rm SU(2)\times U(1)$ subgroup of $\rm SU(3)$. 
We have found that $\mathbb{CP}^2$ can be written as a coset space $\rm SU(3)/(SU(2)\times U(1))$. 
This is apparently a symmetric coset space.

%%%%%%%%%%%%%%%%%%%%%%%%%%%%%%%%%%%%%%%
 %%%%%%%%%%%%%%%%%%%%%%%%%%%%%%%%%%%%%%%

\vspace{1cm}

\section{$G$ as a principal $H$-bundle}   \label{principal bundle appendix}

\vspace{5mm}

In this paper, a coset space $G/H$ is used as a compactification manifold $\cal M$ of a higher-dimensional Yang-Mills theory. 
Since we would like to define a gauge theory on $G/H$ with the gauge group $G_{\rm YM}$, we need a principal $G_{\rm YM}$-bundle on $G/H$ on which we can define the gauge field as a connection. 
Note that, in this paper, we discuss only classical aspects of Yang-Mills theory on a fixed principal $G_{\rm YM}$-bundle, and the summation over different bundles will not be considered.

It is known \cite{Kobayashi-Nomizu} that there exists a natural principal $H$-bundle on $G/H$, as will be reviewed shortly. 
In fact, it is $G$ itself. 
This principal $H$-bundle is specified by a set of transition functions ${\rm g}_{ij}(y)$, whose values are in $H$, defined on the overlap $U_i\cap U_j$ of two local coordinate patches $U_i, U_j\subset G/H$. 
If we choose an embedding of $H$ into the gauge group $G_{\rm YM}$, then ${\rm g}_{ij}(y)$ can be regarded as a set of transition functions whose values are in $G_{\rm YM}$. 
Using these transition functions, we can construct a principal $G_{\rm YM}$-bundle on $G/H$. 
In this manner, we can define Yang-Mills theory on $G/H$ whose gauge group is $G_{\rm YM}$. 

%\vspace{5mm}

Recall the decomposition (\ref{coset decomposition}) of $G$. 
This suggests that $G$ is a fiber bundle whose fibers are of the form $gH$. 
The subset $gH$ is in one-to-one correspondence to $H$ for any $g$. 
In fact, it can be shown that $G$ can be regarded as a fiber bundle whose fibers are $H$, that is, a principal $H$-bundle on $G/H$. 
For the coset spaces, the representatives $g_i$ are replaced with $g(y)$ chosen for each local coordinate patch. 
A choice of $g(y)$ on $U$ amounts to choosing a local section $g(y)$ on $U$ of the principal $H$-bundle $G$. 
The situation is depicted in Figure \ref{G as H-bundle}. 

\begin{figure}
\begin{center}
\includegraphics[width=7.5cm,clip]{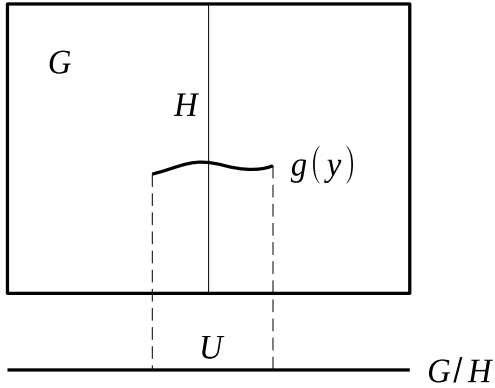}
\end{center}
\caption{
A Lie group $G$ can be regarded as a principal $H$-bundle on $G/H$. 
The vertical direction corresponds to the fiber $H$. 
A local section $g(y)$ on $U$ is also depicted. 
}   \label{G as H-bundle}
\end{figure}

Let $g'(y)$ be another local section on $U$. 
This is related to the original one by 
\begin{equation}
g'(y)\ =\ g(y)h(y), \hspace{1cm} h(y)\in H. 
\end{equation}
Note that this transformation does not transform a point in $U$ to a different one since  
\begin{equation}
g'(y)H\ =\ g(y)h(y)H\ =\ g(y)H. 
\end{equation}
Indeed, this amounts to a local gauge transformation with respect to $h(y)\in H\subset G_{\rm YM}$, as we will show in Appendix \ref{Maurer-Cartan appendix}. 
In fact, this also induces a local Lorentz transformation simultaneously. 
To avoid possible confusions with the ordinary gauge transformation, we call the local transformation induced by the right-multiplication of $h(y)$ an $H$-transformation. 

We can multiply any $y$-independent element $g_0$ to the representatives $g(y)$ from the left. 
The result $g_0\cdot g(y)$ is also an element of $G$, so this must be in a certain subset $g(y')H$ for some point $y'$. 
This can be written as 
\begin{equation}
g_0\cdot g(y)\ =\ g(y')h(y,g_0). 
\end{equation}
In general, the pont $y'$ is different from the original $y$. 
Therefore, the left-multiplication of $g_0$ induces a coordinate transformation of $G/H$. 
In addition, this simultaneously induces an $H$-transformation with respect to $h(y,g_0)$. 
We will see in the next section that this is actually an isometry of $G/H$ for a natural choice of the metric on $G/H$. 
In fact, the isometry group of $G/H$ with respect to the metric is known to be isomorphic to $G$, that is, any isometry of $G/H$ is induced by the left-multiplication as above.

%%%%%%%%%%%%%%%%%%%%%%%%%%%%%%%%%%%%%%%
 %%%%%%%%%%%%%%%%%%%%%%%%%%%%%%%%%%%%%%%
%%%%%%%%%%%%%%%%%%%%%%%%%%%%%%%%%%%%%%%
 %%%%%%%%%%%%%%%%%%%%%%%%%%%%%%%%%%%%%%%

\vspace{1cm}

\section{Maurer-Cartan 1-form}   \label{Maurer-Cartan appendix}

\vspace{5mm}

There exists a 1-form on $G$ which can be defined without any additional information. 
We review in the following that this 1-form defines a natural metric and background gauge field on $G/H$.  
See \cite{Kapetanakis:1992hf} and \cite{Kobayashi-Nomizu} for details. 

Let $g_{ij}(\xi)$ be a matrix-valued function on a local coordinate patch of $G$ whose value at a point $\xi\in G$ is an $N\times N$ matrix representing $\xi$. 
This can be regarded as a set of coordinate functions on $G$. 
Therefore, like $dy^\alpha$ on $U\subset G/H$, we have a set of functions $dg_{ij}(\xi)$ on $G$ whose values are 1-forms on $G$. 
Using them, we construct 
\begin{equation}
(g^{-1})_{ij}(\xi)\,dg_{jk}(\xi). 
\end{equation}
This is called the Maurer-Cartan 1-form on $G$. 
In the following, we simply denote this by $g^{-1}dg$. 

A local section $g(y)$ can be regarded as an embedding of the local coordinate patch $U$ into $G$. 
The pull-back of $g^{-1}dg$ with respect to this embedding gives us a 1-form $g(y)^{-1}dg(y)$ on $U$. 
The latter can be expanded as 
\begin{equation}
g(y)^{-1}dg(y)\ =\ ie^mt_m+ie^at_a, 
\end{equation}
where 
\begin{equation}
e^m\ =\ e^m_\alpha dy^\alpha, \hspace{1cm} e^a\ =\ e^a_\alpha dy^\alpha, 
\end{equation}
since $g(y)^{-1}\partial_\alpha g(y)$ is in the Lie algebra $\mathfrak{g}$. 
We will clarify the meanings of $e^m_\alpha$ and $e^a_\alpha$ in the following. 

Recall that elements of the Lie algebra $\mathfrak{g}$ correspond to tangent vectors of $G$ at the identity element. 
For the coset space, $t_m$ correspond to a set of basis of the tangent space of $G/H$. 
Therefore, $e^m_\alpha$ gives us a vielbein on $G/H$. 
This is an invertible $d\times d$ matrix by construction. 

To see the role played by $e^a_\alpha$, consider an $H$-transformation $g(y)\to g(y)h(y)$. 
This induces 
\begin{equation}
g(y)^{-1}dg(y)\ \to\ h(y)^{-1}\Bigl( g(y)^{-1}dg(y) \Bigr) h(y)+h(y)^{-1}dh(y). 
\end{equation}
This implies 
\begin{equation}
e^m\ \to\ h(y)^{-1}e^mh(y), \hspace{1cm} e^a\ \to\ h(y)^{-1}e^ah(y)-ih(y)^{-1}dh(y). 
\end{equation}
Here, we have used the fact that 
\begin{equation}
h(y)^{-1}t_mh(y)\ \in\ \mathfrak{m}
   \label{induced local Lorentz}
\end{equation}
holds, where $\mathfrak{m}$ is a vector space spanned by $t_m$. 
This is due to the commutation relation 
\begin{equation}
[t_a,t_m]\ =\ if^n{}_{am}t_n
   \label{R_t appendix}
\end{equation}
which we assumed in Appendix \ref{coset space appendix}. 
The transformation of $e^a$ shows that $e^a_\alpha$ behaves as a gauge field on $G/H$ with the gauge group $H$, defined on the principal $H$-bundle $G$. 

Note that $e^m$ also transforms under the $H$-transformation. 
Indeed, the equation (\ref{induced local Lorentz}) implies that this is a local Lorentz transformation acting on the tangent space of $G/H$. 
Therefore, the right-multiplication of $h(y)$ to $g(y)$ induces both a gauge transformation of $e^a_\alpha$ and a local Lorentz transformation of $e^m_\alpha$ simultaneously. 

Recall that the left-multiplication of $g_0\in G$ to a local section $g(y)$ induces a coordinate transformation $y\to y'$ of $G/H$. 
The 1-form $g(y)^{-1}dg(y)$ on $U$ is trivially invariant under $g(y)\to g_0\cdot g(y)$ for a $y$-independent $g_0$. 
This implies that the vielbein $e^m_\alpha$ (and the gauge field $e^a_\alpha$) are invariant under the coordinate transformation. 
In other words, the coordinate transformation is an isometry with respect to the metric $h_{\alpha\beta}$ constructed from $e^m_\alpha$. 
It is known that any isometry of $G/H$ is induced in this manner. 
Therefore, the isometry group of $G/H$ is isomorphic to $G$. 

%\vspace{5mm}

We calculate the field strength 2-form 
\begin{equation}
\frac12f^at_a\ :=\ de^at_a+ie^at_a\wedge e^bt_b\ =\ de^at_a - \frac{1}{2} t_c f^c{}_{ab}  e^a \wedge e^b.
\end{equation}
For this purpose, we do not need to know the explicit form of $e^a_\alpha$. 
Instead, we start with the following identity 
\begin{equation}
d\Bigl( g(y)^{-1}dg(y) \Bigr)\ =\ -g(y)^{-1}dg(y)\wedge g(y)^{-1}dg(y). 
\end{equation}
This can be decomposed into the following two equations 
\begin{eqnarray}
de^at_a &=& \frac12t_cf^c{}_{ab}e^a\wedge e^b+\frac12t_af^a{}_{mn}e^m\wedge e^n, 
   \label{structure eqs 1} \\ [2mm] 
de^mt_m &=& t_mf^m{}_{an}e^a\wedge e^n. 
\label{structure eqs 2}
\end{eqnarray}
The first equation tells us that the field strength is given as 
\begin{equation}
f^a_{\alpha\beta}\ =\ f^a{}_{mn}e^m_\alpha e^n_\beta. 
\end{equation}
On the other hand, the second equation gives us the spin connection 
\begin{equation}
\omega_\alpha{}^m{}_n\ =\ -f^m{}_{an}e^a_\alpha. 
\end{equation}
Then, the curvature 
\begin{equation}
\frac12R_{\alpha\beta}{}^m{}_ndy^\alpha\wedge dy^\beta\ :=\ d\omega^m{}_n+\omega^m{}_k\wedge \omega^k{}_n
\end{equation}
can be also calculated explicitly. 
We obtain 
\begin{equation}
R_{\alpha\beta}{}^m{}_n\ =\ -f^a{}_{kl}f^m{}_{an}e^k_\alpha e^l_\beta. 
   \label{curvature appendix}
\end{equation}

%%%%%%%%%%%%%%%%%%%%%%%%%%%%%%%%%%%%%%%
 %%%%%%%%%%%%%%%%%%%%%%%%%%%%%%%%%%%%%%%

\vspace{1cm}

\section{Equations of motion for the background flux}
\label{EOM appendix}
\vspace{5mm}

In this appendix, we show that the background gauge field $\bar{A}_\alpha$ defined in section \ref{{Metric and background gauge field on $G/H$ }} satisfies the equations of motion with respect to the vielbein defined in Appendix \ref{Maurer-Cartan appendix}. 

The equations of motion are 
\begin{equation}
\nabla^\alpha\bar{F}_{\alpha\beta}+i[\bar{A}^\alpha,\bar{F}_{\alpha\beta}]\ =\ 0, 
   \label{equations of motion appendix}
\end{equation}
where 
\begin{equation}
\bar{A}_\alpha\ =\ e^a_\alpha T_a, \hspace{1cm} \bar{F}_{\alpha\beta}\ =\ e^m_\alpha e^n_\beta f^a{}_{mn}T_a. 
\end{equation}
The covariant derivative is defined with respect to the spin connection as 
\begin{eqnarray}
\nabla^\alpha\bar{F}_{\alpha\beta} 
&=& e^n_\beta\left( \partial^m\bar{F}_{mn}-\omega^{ml}{}_m\bar{F}_{ln}-\omega^{ml}{}_n\bar{F}_{ml} \right) \nonumber \\ [2mm]
&=& e^n_\beta e^{m\alpha}e^a_\alpha\left( f^l{}_{am}f^b{}_{ln}+f^l{}_{an}f^b{}_{ml} \right)T_b.
\end{eqnarray}
The commutator can be written as 
\begin{equation}
i[\bar{A}^\alpha,\bar{F}_{\alpha\beta}]\ =\ -e^n_\beta e^{m\alpha}e^a_\alpha f^b{}_{ac}f^c{}_{mn}T_b. 
\end{equation}
By using the Jaocbi identity and the fact that $G/H$ is symmetric, we find that the equations of motion (\ref{equations of motion appendix}) are automotaically satisfied. 

%%%%%%%%%%%%%%%%%%%%%%%%%%%%%%%%%%%%%%%%%%%%%%%%%%%%%%
%%%%%%%%%%%%%%%%%%%%%%%%%%%%%%%%%%%%%%%
 %%%%%%%%%%%%%%%%%%%%%%%%%%%%%%%%%%%%%%%

\vspace{1cm}

\section{Explicit constructions of backgrounds in the $\rm{SU}(2)/{\rm U}(1)$ coset space}   \label{app: explicit formulas}

\vspace{5mm}

In this appendix, we explicitly calculate $e^m_\alpha$ and $e^a_\alpha$ defined in Appendix \ref{Maurer-Cartan appendix} for the case $G/H=S^2$. 
We will see that $e^m_\alpha$ is the standard zweibein which gives the round metric on $S^2$, and $e^a_\alpha$ describes a monopole configuration on $S^2$. 

Any element $g$ of $\rm SU(2)$ can be written as 
\begin{equation}
g\ =\ \left[
\begin{array}{cc}
a & -b^* \\
b & a^*
\end{array}
\right], \hspace{1cm} |a|^2+|b|^2\ =\ 1. 
\end{equation}
The Maurer-Cartan 1-form $g^{-1}dg$ is then given as 
\begin{equation}
g^{-1}dg\ =\ \sigma_+(b^*da^*-a^*db^*)+\sigma_-(adb-bda)+\sigma_3(a^*da+b^*db), 
\end{equation}
where $\sigma_\pm$ and $\sigma_3$ are the Pauli matrices. 

We choose a local section $g(\theta,\varphi)$ by restricting $a,b$ to be 
\begin{equation}
a\ =\ \cos\frac\theta2, \hspace{1cm} b\ =\ e^{i\varphi}\sin\frac\theta2, 
\end{equation}
where $\theta,\varphi$ are the polar and the azimuthal angles of $S^2$, respectively. 
The pull-back of $g^{-1}dg$ by this local section is then 
\begin{equation}
g(\theta,\varphi)^{-1}dg(\theta,\varphi)\ =\ ie^+t_++ie^-t_-+ie^3t_3, 
\end{equation}
where 
\begin{equation}
t_\pm\ :=\ \sigma_\pm, \hspace{1cm} t_3\ :=\ \frac12\sigma_3, 
\end{equation}
and 
\begin{equation}
e^\pm\ :=\ \pm\frac i2e^{\mp i\varphi}(d\theta\mp i\sin\theta d\varphi), \hspace{1cm} e^3\ := (1-\cos\theta)d\varphi. 
\end{equation}
The metric $h_{\alpha\beta}$ obtained from $e^\pm$ is therefore the round metric 
\begin{equation}
ds^2\ =\ 4e^+e^-\ =\ d\theta^2+\sin^2\theta d\varphi^2, 
\end{equation}
as expected. 
The field strength obtained from $e^3$ is 
\begin{equation}
de^3\ =\ \sin\theta\,d\theta d\varphi. 
\end{equation}
The integral of this 2-form gives 
\begin{equation}
\frac1{2\pi}\int_{S^2}\sin\theta\,d\theta d\varphi\ =\ 2. 
\end{equation}
This shows that $e^3_\alpha$ describes a $\rm U(1)$ monopole configuration with the monopole charge $2$. 

Note that the radius of $S^2$ is set to be $1$ in the above expressions. 
It is rather easy to recover the radius $a$ based on the dimensional analysis since $a$ is essentially the only dimensionful parameter in the Kaluza-Klein reduction of Yang-Mills theory. 
In higher dimensions, the coupling constant $g_{\rm YM}$ is dimensionful, but it is just an overall coefficient in the action. 

%\vspace{5mm}

In the following, we present an explicit description of the model of $S^2$=SU$(2)/$U$(1)$ compactification.
We will employ different convensions from those in the main body of this paper, which might be more familiar to the readers. 

The action of the six-dimensional space-time is
\begin{equation}
 S = \int d^6x \sqrt{-g}
  \left\{
   \frac{1}{\kappa^2}{\cal R} -\frac{1}{2g^2} {\rm Tr} \left( F^{MN} F_{MN} \right) - \Lambda
  \right\},
\label{app:action}
\end{equation}
 where ${\cal R}$ and $\Lambda$ are Ricci scalar and cosmological constant, respectively, and
\begin{equation}
 F_{MN} = \nabla_M A_N - \nabla_N A_M - i [A_M,A_N]
\end{equation}
 is the gauge field strength with the covariant derivative
\begin{equation}
 \nabla_M A_N = \partial_M A_N - \Gamma^L{}_{MN} A_L
\end{equation}
 and $M,N = 0,1,2,\cdots,4,5$.
The field strength is matrix valued as $F_{MN} = T^a F^a_{MN}$,
 where $T^a$ is the generator matrix of the gauge group $G_{\rm YM}=$\, SU$(3)$ with
\begin{equation}
 {\rm Tr}(T^a T^b) = \frac{1}{2}\delta^{ab},
 \qquad
 [T^a,T^b] = i f^{abc}T^c.
\end{equation}
The metric is described as
\begin{equation}
 ds^2 = - dt^2 + \sum_{i=1,2,3} dx^i dx^i + a^2 \left( d\theta^2 + \sin^2 \theta d\varphi^2 \right).
\end{equation}
The sphere of radius $a$ is described by two angular coordinates, $\theta$ and $\varphi$,
 and non-compact spacetime is the Minkowski space.
A vector on the sphere is described by two independent basis vactors,
 ${\bf e}_\theta$ and ${\bf e}_\varphi$,
 which correspond to the two of three unit vectors
 of the polar coordinate system of three dimensional flat space.

Introduce a background gauge field on the sphere
\begin{equation}
 {\bar {\bf A}} = {\bar A}_\theta {\bf e}_\theta + {\bar A}_\varphi {\bf e}_\varphi,
\qquad
 {\bar A}_\theta = 0,
\qquad
 {\bar A}_\varphi = \frac{1}{2} H \frac{\cos\theta \mp 1}{a \sin\theta},
\label{app:back}
\end{equation}
 where negative sign is for $0 < \theta < \pi/2$ and
 positive sign is for $\pi/2 < \theta < \pi$ in $A_\varphi$.
Note that ${\bar A}_4 = a {\bar A}_\theta$ and ${\bar A}_5 = a \sin\theta {\bar A}_\varphi$,
 since $x_4=\theta$ and $x_5=\varphi$.
This is a spherical slice of monopole configuration with unit magnetic charge at radius $a$,
 and the generator of corresponding U$(1)$ gauge symmetry is $H/2$.
We can explicitly write
\begin{equation}
 H = \left(
      \begin{array}{ccc}
       n_1 & 0 & 0 \\
      0  & n_2 & 0 \\
       0 & 0 & -n_1-n_2 
      \end{array}
     \right).
\end{equation}
The off-diagonal components of the matrix-valued SU$(3)$ gauge field
 transform as matter fields under this U$(1)$ gauge symmetry with charges
\begin{equation}
 \left(
  \begin{array}{ccc}
   * & (n_1-n_2)/2 & (2n_1+n_2)/2 \\
   -(n_1-n_2)/2    & * & (n_1+2n_2)/2\\
   -(2n_1+n_2)/2   & -(n_1+2n_2)/2 & * 
  \end{array}
 \right),
\end{equation}
 and the Dirac quantization of electric charge by a monopole indicates
 that twice of each charge should be an integer, which indicates
\begin{equation}
 n_1-n_2 \in {\bf Z},
\qquad
 3n_1 \in {\bf Z},
\qquad
 3n_2 \in {\bf Z}.
\end{equation}
This background configuration
 is invariant under the symmetry transformation of $S^2$, or $\rm SU(2)$,
 up to corresponding gauge transformation generated by $H/2$.
Furthermore, 
 we can make this background configuration a solution of field equations
 of the action, (\ref{app:action}), namely SU$(3)$ Yang-Mills equation
 and Einstein equation by choosing
 $\Lambda = 1/\kappa^2a^2$ and $g^2={\rm Tr}((H/2)^2)\kappa^4\Lambda$ \cite{Randjbar-Daemi:1982bjy}.

The fluctuation around the background configuration of (\ref{app:back}), $\delta A^4$ and $\delta A^5$,
 can be described as scalar fields in low-energy four-dimensional effective theory.
It is convenient to describe these field 
 so that they are living on the tangent space of $S^2$.
Introduce zweibein as
\begin{equation}
 g_{\mu\nu} = e_\mu{}^m \, \delta_{mn} \, e^n{}_\nu,
\end{equation}
 where $\mu,\nu=4,5$, $m,n=4,5$ and in the matrix form $g_{\mu\nu}=a^2{\rm diag}(1, \sin^2\theta)$.
Explicitly we specially introduce
\begin{equation}
 e^n{}_\nu = a \left(
                \begin{array}{cc}
                 \cos\varphi & -\sin\varphi \\
                 \sin\varphi & \cos\varphi
                \end{array}
               \right)
               \left(
                \begin{array}{cc}
                 1 & 0 \\
                 0 & \sin\theta
                \end{array}
               \right)
\end{equation}
 which satisfies the above formula of definition.
The fields on tangent space is defined as
\begin{equation}
 V^n \equiv e^n{}_\nu \, \delta A^\nu .
\end{equation}
Furthermore, it is convenient to define a complex scalar field as
\begin{equation}
 V_{\pm} \equiv \frac{1}{\sqrt{2}} \left( V^4 \mp i V^5 \right),
\qquad
 V_- = (V_+)^\dag .
\end{equation}
We define a covariant derivative with background field as
\begin{equation}
 {\bar D}_\mu V_{\pm} \equiv \nabla_\mu V_\pm - i [ {\bar A}_\mu, V_\pm ].
\end{equation}
The explicit form can be obtained as
\begin{equation}
 {\bar D}_\mu V_{\pm}
  = \partial_\mu V_\pm
    -i \, \delta^5_\mu \, (\cos\theta - 1) \left\{ \pm V_\pm +\left[ \frac{H}{2},V_\pm \right] \right\}
\end{equation}
 for $0 < \theta < \pi/2$.
In case of $\pi/2 < \theta < \pi$,
 the factor $(\cos\theta-1)$ in the second term should be replaced by $(\cos\theta+1)$.
Note that the field $V_\pm$ is matrix valued as $V_\pm^a T^a$
 and the second term can vanish depending on the choice of $H$.

%%%%%%%%%%%%%%%%%%%%%%%%%%%%%%%%%%%%%%%%%%%%%%%%%%%%%%
%%%%%%%%%%%%%%%%%%%%%%%%%%%%%%%%%%%%%%%%%%%%%%%%%%%%%%
%%%%%%%%%%%%%%%%%%%%%%%%%%%%%%%%%%%%%%%%%%%%%%%%%%%%%%
%%%%%%%%%%%%%%%%%%%%%%%%%%%%%%%%%%%%%%%%%%%%%%%%%%%%%%

\vspace{1cm}

\section{Mode expansions  on $G/H$}   \label{mode functions appendix}

\vspace{5mm}

In order to perform the Kaluza-Klein reduction on $G/H$, we need to have a complete set of functions on $G/H$ by which any fields can be expanded. 
More precisely, we need to expand sections of suitable vector bundles on $G/H$, not only functions, since there are fields with a local Lorentz indices belonging to a non-trivial representation of the gauge group. 
For $S^2$, the monopole harmonics \cite{Wu:1976ge} play such a role. 
In more general cases, it is known that there exists a useful set of functions on $G$ \cite{Salam:1981xd} which can be employed for our purpose, as we will review in this appendix. 

\vspace{5mm}

{\bf Mode expansions on $G$} \\
It is rather easy to find examples of functions on $G$. 
Let $\rho:G\to {\rm GL}(n,\mathbb{C})$ be an $n$-dimensional representation of $G$. 
This assigns to each element $g$ of $G$ an $n\times n$ matrix $\rho(g)$. 
Each matrix component $\rho(g)_{IJ}$ is therefore a function on $G$. 
For each $n$, or more appropriately, for each representation $R$ of $G$, we can obtain many functions on $G$ in this manner. 
It can be shown that they are linearly independent. 

The Peter-Weyl theorem tells us that any function on $G$ can be expanded by these functions. 
Explicitly, a function $f(g)$ on $G$ can be written as 
\begin{equation}
f(g)\ =\ \sum_R\sum_{I,J=1}^{d_R}\tilde{c}^R_{IJ}\,\rho^R(g)_{IJ}, 
\end{equation}
where the first sum is taken over all the representations of $G$ with the multiplicity one, $d_R$ is the dimension of the representation $R$, and $\rho^R(g)$ is the matrix representing $g\in G$ on the representation $R$. 
For a later purpose, we employ an equivalent expansion 
\begin{equation}
f(g)\ =\ \sum_R\sum_{I,J=1}^{d_R}c^R_{IJ}\,\rho^R(g^{-1})_{IJ}. 
   \label{mode expansion appendix}
\end{equation}
This is also valid since any function $f(g)$ on $G$ can be written as $\tilde{f}(g^{-1})$ by using some $\tilde{f}(g)$.

\vspace{5mm}

{\bf Mode expansions of scalar functions on $G/H$} \\
We then study a complete set of functions on $G/H$ by using its extension  to $G$ as follows. 
Let $\phi(y)$ be a function on  a local coordinate patch $U$ of $G/H$. 
By using a local section $g(y) \in G$, we can regard $\phi(y)$ as a function $\Phi(g(y))$ defined only on a subset of $G$ which is the embedding of $U$. 
We can extend $\Phi(g(y))$ to the fiber direction by an action $h(y) \in H$ as 
\begin{equation}
\Phi(g(y)h(y))\ =\ \phi(y), \hspace{1cm} h(y)\in H. 
   \label{extension of function appendix}
\end{equation}
In this manner, any function $\phi(y)$ on $G/H$ can be extended to a function $\Phi(g)$ on $G$. 
Thus a complete set of functions on $G/H$ can be obtained from a complete set of functions on $G$
by imposing the condition 
\begin{equation}
{\Phi(g(y)h(y))\ =\ \Phi(g(y)). }
\end{equation}
Namely, these functions must take 
constant values along the fiber direction {corresponding to} $H$-transformations. 

Any function on $G$ can be uniquely expanded as (\ref{mode expansion appendix}). 
The representation $R$ of $G$ is decomposed into various irreducible representations of $H$, 
 which may contain unit representation ${\bf 1}$. 
A complete set of  scalar functions $\phi(y)$ on $G/H$ is then given by a complete set of constant functions
along the fiber $H$, which correspond to 
the unit representations constructed from all the representations $R$ of $G$. 
An explicit form of the expansion is discussed as a special case of  non-scalar functions (sections)
which transform nontrivially as representation $r$ of $H$. 

\vspace{5mm}

{\bf Mode expansions of $a_\mu$ and $\phi_m$ on $G/H$} \\
In the Kaluza-Klein reduction of Yang-Mills theory, we would llike to expand $a_\mu$ and $\phi_m$. 
Since they are sections of some vector bundles on $G/H$ and transforms nontrivially under $H$, 
we need to modify the above expansion procedure as follows. 

First, consider $a_\mu$. 
This belongs to the adjoint representation $\rm ad_{YM}$ of $G_{\rm YM}$. 
Recall that, as explained in Appendix \ref{principal bundle appendix}, our principal $G_{\rm YM}$-bundle is constructed from the principal $H$-bundle $G$, and the transition functions take values in a subgroup $H$ of $G_{\rm YM}$. 
This means that we can consider $a_\mu$ as a field defined on the principal $H$-bundle $G$. 
Since $\rm ad_{YM}$ is reducible for $H$, we decompose $a_\mu$ into components according to the irreducible decomposition of $\rm ad_{YM}$ with respect to $H$. 
Each component belongs to an irreducible representation of $H$, and forms a section of a vector bundle on $G/H$. 
Note that this decomposition is compatible with the gauge symmetry preserved by the background flux $\bar{F}_{\alpha\beta}$ since the preserved symmetry corresponds to a subgroup of $G_{\rm YM}$ which commutes with $H$. 

The case for $\phi_m$ requires one more twist since it has the tangent space index $m$, and 
transforms under the local Lorentz transformations. 
As explained under (\ref{covariant derivative general}),
 the covariant derivative $\nabla_\alpha$ with respect to the metric
 on $G/H$ can be identified with the one with respect to the background gauge field
 in the representation $R_t$ of $H$, which is given by the commutation relation (\ref{R_t appendix}).
Thus,  the tangent space indices are regarded as indices of the representation $R_t$ of $H$. 
Then, $\phi_m$ belongs to a product representation $R_t\otimes{\rm ad_{YM}}$ of $H$. 
We decompose $\phi_m$ into components according to the irreducible decomposition of $R_t\otimes {\rm ad_{YM}}$ with respect to $H$. 
Each component again forms a section of a vector bundle on $G/H$. 

To summerize, the mode expansions of $a_\mu$ and $\phi_m$ can be performed if we know how to expand a section of a vector bundle on 
$G/H$ which belongs to an irreducible representation of $H$. The mode expansions of $a_\mu$ or $\phi_m$
are given by a sum of various mode functions corresponding to each irreducible representations of $H$.

\vspace{5mm}

{\bf Mode expansions of $\chi_i(y)$ in representation $r$ of $H$} \\
Now, we consider the expansion of $\chi_i(y)$ which belongs to a specific representation $r$ of $H$. 
Recall a local section $g(y)$ on $G/H$ can be extended to the {fiber direction} by the  $H$-transformation as
$$
g(y)\to g(y)h(y). 
$$
Thus, any $g \in G$ is written as a product of the local section $g(y)$ and the $H$-transformation $h(y)$. 
This induces an $H$-transformation on $\chi_i(y)$ as 
$$ 
\chi_i(y)\to\rho^r(h(y)^{-1})_{ij}\chi_j(y).
$$
Thus, we can extend the function $\chi_i(y)$ on $G/H$ to a function ${\cal X}_i(g)$ on $G$ by
\begin{equation}
 {\cal X}_i(g)\ := \ 
{\cal X}_i(g(y)h(y))\ =\ \rho^r(h(y)^{-1})_{ij}\chi_j(y),
\end{equation}
where $g=g(y)h(y)$. 
If $r$ is the trivial representation of $H$, then this reduces to (\ref{extension of function appendix}). 
By construction, ${\cal X}_i(g)$ satisfies 
\begin{equation}
{\cal X}_i(gh)\ =\ \rho^r(h^{-1})_{ij}{\cal X}_j(g). 
   \label{condition for section appendix}
\end{equation}

%\vspace{5mm}

Since each component  ${\cal X}_i(g)$ 
is a function on $G$, it is expanded as (\ref{mode expansion appendix}). 
Then, the above condition (\ref{condition for section appendix})
 imposes the following restriction on 
 the expansion coefficients $c^R_{IJ}$, namely an allowed set of functions on $G$. 
For a representation $R$ of $G$, 
the function $\rho^R(g^{-1})_{IJ}$ on $G$ satisfies the transformation law
\begin{equation}
\rho^R((gh)^{-1})_{IJ}\ =\ \rho^R(h^{-1}g^{-1})_{IJ}\ =\ \rho^R(h^{-1})_{IK}\rho^R(g^{-1})_{KJ}
\end{equation}
under an action of $h \in H$.
This representation $R$ of $G$
can be decomposed into irreducible representations $r_1\oplus \cdots \oplus r_l$ of $H$. 
For the basis according to this decomposition, the matrix $\rho^R(h^{-1})_{IK}$ takes a block-diagonal form. 
Let $i_1,\cdots, i_l$ be indices corresponding to the representations $r_1,\cdots,r_l$, respectively. 
Then, $\rho^R(g^{-1})_{i_1K}, \cdots, \rho^R(g^{-1})_{i_lK}$ transform separately as 
\begin{equation}
\rho^R((gh)^{-1})_{i_a J}\ =\ \rho^{r_a}(h^{-1})_{i_aj_a}\rho^R(g^{-1})_{j_aJ}, \hspace{1cm} (a=1,\cdots,l)
\label{r-transformation under H}
\end{equation}
under the $H$-transformation. 
Let us first consider the case when the representation $r_1$ is isomorphic to $r$. 
Then a rectangular part $\rho^R(g^{-1})_{i_1J}$ satisfies the condition (\ref{condition for section appendix}). 
In general, the decomposition of $R$ contains a multiple of $r$-representation of $H$.
Suppose that $r_a=r$ for $a=1,\cdots,k$ among $(r_1, \cdots r_l).$
We denote the corresponding rectangular part by $\rho^{R,a}(g^{-1})_{iJ}$ where $i$ is the index for $r$. 
Then, the function ${\cal X}_i(g)$ in the representation $r$ can be expanded in terms of these functions as
\begin{equation}
{\cal X}_i(g)\ =\ \sum_{r\subset R}\sum_{J=1}^{d_R}\sum_{a=1}^k c^{R,a}_J\,\rho^{R,a}(g^{-1})_{iJ}. 
\end{equation}
The first sum is taken over all the representations $R$ of $G$ whose irreducible decomposition with respect to $H$ contains $r$. 
Finally, the expansion of $\chi_i(y)$ is given as 
\begin{equation}
\chi_i(y)\ =\ \sum_{r\subset R}\sum_{J=1}^{d_R}\sum_{a=1}^kc^{R,a}_{J}\,f^{R,a}_{iJ}(y), 
\end{equation}
where 
\begin{equation}
 f^{R,a}_{iJ}(y)\ :=\ \rho^{R,a}(g(y)^{-1})_{iJ}. 
\end{equation}
These functions satisfy the same transformation law of (\ref{r-transformation under H}) where $r_a=r$. 
In this paper, we sometimes suppress the index $a$ for notational simplicity.

%%%%%%%%%%%%%%%%%%%%%%%%%%%%%%%%%%%%%%%%%%%%%%%%%%%%%%
%%%%%%%%%%%%%%%%%%%%%%%%%%%%%%%%%%%%%%%%%%%%%%%%%%%%%%
%%%%%%%%%%%%%%%%%%%%%%%%%%%%%%%%%%%%%%%%%%%%%%%%%%%%%%
%%%%%%%%%%%%%%%%%%%%%%%%%%%%%%%%%%%%%%%%%%%%%%%%%%%%%%

\vspace{1cm}

\section{Laplacian and the mass formula on $G/H$}
\label{app: laplacian and mass formula}
\vspace{5mm}

In the Kaluza-Klein reduction,
 the mass of each mode is typically related to
 the eigenvalue of the Laplacian of the compactification manifold. 
The eigenvalues of the Laplacian on coset spaces were discussed in \cite{Pilch:1984xx}.
Interestingly, the mode functions we introduced in Appendix \ref{mode functions appendix}
 turn out to be the eigenfunctions of the Laplacian on $G/H$ provided that
 the vielbein $e^m_\alpha$ and the background gauge field $\bar{A}_\alpha$
 are given as in section \ref{sec:KKcoset} and Appendix \ref{Maurer-Cartan appendix}
 \cite{Palla:1983re,Schellekens:1984dm}.

%\vspace{5mm}

First, we show that the action of the covariant derivative $\bar{D}_\alpha$ on the mode functions $f^R_{IJ}(y)$ 
can be written in an algebraic form. 
Since $f^R_{IJ}(y)$ can be written as $\left( \rho^R(g(y))^{-1} \right)_{IJ}$, the exterior derivative is given as 
\begin{equation}
d\rho^R(g(y))^{-1}\ =\ -\rho^R(g(y))^{-1}d\rho^R(g(y))\cdot \rho^R(g(y))^{-1}, 
   \label{derivative inverse appendix}
\end{equation}
where the matrix indices are suppressed. 
The right-hand side contains the pull-back of the Maurer-Cartan 1-form in the representation $R$. 
This can be expanded as 
\begin{equation}
\rho^R(g(y))^{-1}d\rho^R(g(y))\ =\ ie^mT^R_m+ie^aT^R_a, 
\end{equation}
where $T^R_m, T^R_a$ are the generators of $\mathfrak{g}$ in the representation $R$. 
Inserting this expression into (\ref{derivative inverse appendix}), we obtain 
\begin{equation}
d\rho^R(g(y))^{-1}+ie^aT^R_a\rho^R(g(y))^{-1}\ =\ -ie^mT^R_m\rho^R(g(y))^{-1}. 
\label{covariant derivative algebraic}
\end{equation}
Since the background gauge field is given by $\bar{A}=e^a_\alpha T_a dy^\alpha$, 
this can be written as 
\begin{equation}
\bar{D}_\alpha f^R_{IJ}(y)\ =\ -ie^m_\alpha (T^R_m)_{IK}f^R_{KJ}(y). 
\end{equation}
Thus, the group theoretic argument shows that 
the covariant derivative  $\bar{D}_\alpha$ on $f^R_{IJ}(y)$ can be simply written as a multiplication of 
 $-ie^m_\alpha T^R_m$. 
Note that the covariant derivative $\bar{D}_\alpha$ originally contains the spin connection on $G/H$, 
but as explained below (\ref{covariant derivative general}), we can regard the local Lorentz indices 
as a charge of the gauge group $H$ on $G/H$. Thus the spin connection term in $\bar{D}_\alpha$ 
is absorbed into the gauge connection. The representation matrices of $H$ are thus given by the tensor product of the 
local Lorentz representation and the original $H$ charge. 

%\vspace{5mm}

In section \ref{condensation  S^2}, in particular in (\ref{covariant derivative generator S^2}), 
 we use this relation for $G={\rm SU}(2)$. 
The representations of ${\rm SU}(2)$ are labeled by the half integers $j$. 
The spin-$j$ representation is give by the matrix-valued function $\rho^{(j)}(g)_{mm'}$ where $-j\le m,m'\le j$. 
We define the mode functions on $G/H$ by
\begin{equation}
f^j_{mm'}(y)\ :=\ \rho^{(j)}(g(y)^{-1})_{mm'} ,
\end{equation}
where $m$ is constrained by  the  $H=U(1)$ charge $q$. The condition is discussed in (\ref{S2-Tcharge-q}). 

The covariant derivative $\bar{D}_+$ acting on 
$f^j_{mm'}(y)$ can be rewritten as a multiplication of  $-i\left( T^{(j)}_+ \right)_{mm'}$ on the function as
 (\ref{covariant derivative generator S^2}).  
Note that the charge quantization discussed in Appendix \ref{app: explicit formulas} is automatically satisfied. 

%\vspace{5mm}

Next, we show that the mode functions $f^R_{iJ}(y)$ are eigenfunctions of the Laplacian $h^{\alpha\beta}\bar{D}_\alpha\bar{D}_\beta$, where $i$ is the index for an irreducible representation $r$ of $H$ which is contained in $R$ \cite{Pilch:1984xx}. 
Indeed, 
\begin{eqnarray}
-h^{\alpha\beta} \bar{D}_\alpha\bar{D}_\beta f^R_{iJ}(y)
&=& -h^{\alpha\beta}\bar{D}_\alpha\left( -ie^n_\beta (T^R_n)_{iK}f^R_{KJ}(y) \right) \nonumber \\ [1mm] 
&=& ih^{\alpha\beta}e^n_\beta\left( (T^R_n)_{iK}\partial_\alpha f^R_{KJ}(y)+e^a_\alpha f^m{}_{an}(T^R_m)_{iK}f^R_{KJ}(y)+ie^a_\alpha (T^R_aT^R_n)_{iK}f^R_{KJ}(y) \right) \nonumber \\ [1mm] 
&=& ih^{\alpha\beta}e^n_\beta (T^R_m)_{iK}\bar{D}_\alpha f^R_{KJ}(y)+ie^a_\alpha\left( -if^m{}_{an}T^R_m+[T^R_a,T^R_n] \right)_{iK}f^R_{KJ}(y) \nonumber \\ [1mm] 
&=& \delta^{mn}(T^R_mT^R_n)_{iK}f^R_{KJ}(y) \nonumber \\ [1mm] 
&=& \left( c^G_2(R)-c^H_2(r) \right)f^R_{iJ}(y), 
\end{eqnarray}
where $c_2^G(R)$ is the second Casimir invariant of the representation $R$ of $G$. 

This eigenvalue gives us the mass of each Kaluza-Klein mode of $a_\mu$. 
Recall that, for the mode expansion of $a_\mu$ reviewed in Appendix \ref{mode functions appendix}, we first need to decompose the adjoint representation ${\rm ad}_{\rm YM}$ of $G_{\rm YM}$ with respect to $H$. 
Let $r$ be one of the irreducible representations of $H$ appearing in the decomposition of $\rm ad_{YM}$. 
The components of $a_\mu$ corresponding to $r$ are expanded in terms of $f^R_{iJ}(y)$ where $R$ is a representation of $G$ whose irreducibe decomposition with respect to $H$ contains $r$. 
This vector mode has the mass given as 
\begin{equation}
m^2_v\ =\ m^2_{R,r}\ :=\ c^G_2(R)-c^H_2(r). 
\end{equation}
Note that $m^2_{R,r}$ is always non-negative. 

Next, we consider the masses of scalar modes obtained from $\phi_m$. 
Recall that the mass terms of the modes come from 
\begin{equation}
{\rm Tr}\left[ \frac12\left( \bar{D}_\alpha\phi_\beta \right)^2-\frac12\phi^\alpha R_{\alpha\beta}\phi^\beta-i\phi^\alpha [\bar{F}_{\alpha\beta},\phi^\beta] \right]. 
\end{equation}
The first term gives $m^2_{R,r}$. 
The explicit expression of the curvature (\ref{curvature appendix}) implies that the Ricci tensor is given as 
\begin{equation}
R_{\alpha\beta}\ =\ -c_2^H(R_t)e^m_\alpha e^n_\beta\,{\rm tr}(t_mt_n), 
\end{equation}
where $t_m$ belong to the representation used to define the coset space $G/H$. 
Therefore, the curvature contribution to the mass is given as 
\begin{equation}
\frac12c_2^H(R_t)^m{}_n\,{\rm Tr}\,\phi_m\phi^n. 
\end{equation}
The flux contribution can be written as 
\begin{equation}
\phi_{mA}(T^{R_t}_a)^m{}_n(T^{a})^A{}_B\phi^{nB}, 
\end{equation}
where $\phi_m=\phi_m^AT_A$ is the expansion of $\phi_m$ in terms of the generators $T_A$ of $G_{\rm YM}$, and $(T_a)^A{}_B$ are the generators of $\mathfrak{h}$ represented on ${\rm ad}_{\rm YM}$ which is reducible with respect to $\mathfrak{h}$. 
Like in the case of the angular momentum in quantum mechanics, this can be written as 
\begin{equation}
\frac12\phi_{mA}\left( c_2^H(R_t\otimes {\rm ad}_{\rm YM})^{mA}{}_{nB}-c_2^H(R_t)^m{}_n\delta^A_B-c_2^H({\rm ad}_{\rm YM})^A{}_B\delta^m_n \right)\phi^{nB}. 
\end{equation}
The second term cancels the curvature contribution. 

In order to determine the mass explicitly, we consider the decomposition of $\phi_m$ in more detail. 
This belongs to $R_t\otimes {\rm ad_{YM}}$. 
First, we decompose $\rm ad_{YM}$ and pick up one irreducible representation $\tilde{r}$ of $H$. 
The corresponding components of $\phi_m$ belong to $R_t\otimes \tilde{r}$. 
We further decompose $R_t\otimes\tilde{r}$ and pick up $r$. 
These components are expanded in terms of $f^R_{iJ}(y)$ where $R$ contains $r$. 
Their masses are therefore given as 
\begin{equation}
m_s^2\ =\ m^2_{R,r}+c_2^H(r)-c_2^H(\tilde{r})\ =\ c_2^G(R)-c_2^H(\tilde{r}). 
\end{equation}

The important difference of this mass formula from $m_v^2$ is that the second term in the right-hand side is the Casimir invariant for $\tilde{r}$, not for $r$. 
For example, even if $\tilde{r}$ is non-trivial, $R_t\otimes \tilde{r}$ may contain a singlet component. 
In this case, we can choose $R={\bf 1}$, so that $m^2$ is negative. 

Indeed, this happens for the symmetric Higgs fields. 
As defined in section \ref{sec:symmetric Higgs}, a symmetric Higgs field consists of those components of $\phi_m$ which is singlet with respect to $H$. 
This means $r={\bf 1}$. 
Therefore, the mode expansion of the symmetric Higgs field do contain the contribution from $R={\bf 1}$. 
This is nothing but the constant mode, and the mass is given as 
\begin{equation}
m^2\ =\ -c_2^H(\tilde{r}). 
   \label{mass of symmetric Higgs appendix}
\end{equation}
As long as $\tilde{r}$ is non-trivial, this mode has a tachyonic mass term which allows the symmetric Higgs field to acquire a non-zero vacuum expectation value. 
This fact justifies their name.

%%%%%%%%%%%%%%%%%%%%%%%%%%%%%%%%%%%%%%%%%%%%%%%%%%%%%%

\vspace{1cm}

\section{Symmetric Higgs fields are symmetric}
\label{app: symmetric Higgs field}
\vspace{5mm}

In this Appendix, we show that the symmetric Higgs fields defined in section \ref{sec:symmetric Higgs}
 are symmetric fields in the sense of \cite{Forgacs:1979zs}, as the name suggests. 

First, we introduce the notion of $G$-invariant fields. 
A field $\phi_m$ on $G/H$ is said to be $G$-invariant if this satisfies 
\begin{equation}
g_0\cdot\phi_m(y)\ =\ \phi_m(y), \hspace{1cm} g_0\in G, 
   \label{G-invariance appendix}
\end{equation}
where the action of $g_0$ is defined as 
\begin{equation}
g_0\cdot\phi_m(y)\ :=\ \Phi_m(g_0g(y)), 
\end{equation}
where $\Phi_m(g)$ are a set of functions on $G$ which extend $\phi_m(y)$ on $G/H$
to the fiber directions $H$, as explained in Appendix \ref{mode functions appendix}.  
Recall that $g_0g(y)$ can be written as $g(y')h(y,g_0)$, where $y\to y'$ is an isometry of $G/H$ induced by $g_0$
and $h(y,g_0) \in H$ is {an $H$-transformation}. 
Then, the condition (\ref{G-invariance appendix}) can be written as 
\begin{equation}
\Lambda_{mn}(y)U(y)\phi_n(y')U(y)^\dag\ =\ \phi_m(y), 
\end{equation}
where $\Lambda_{mn}(y)$ and $U(y)$ are the local Lorentz transformation and the gauge transformation induced by $h(y,g_0)$. 
This shows that a $G$-invariant field is symmetric. 

Recall that a symmetric Higgs field $\phi_m$ is defined to satisfy $\bar{D}_\alpha \phi_n=0$ and $\partial_\alpha \phi_n=0$. 
These imply that $\phi_m$ is $y^\alpha$-independent and $H$-invariant. 
Then, we find 
\begin{equation}
\Lambda_{mn}(y)U(y)\phi_n(y')U(y)^\dag\ =\ \Lambda_{mn}(y)U(y)\phi_n(y)U(y)^\dag\ =\ \phi_m(y). 
\end{equation}
Therefore, $\phi_m$ is $G$-invariant. 
This then implies that it is symmetric.

\end{document}